\documentclass[conference]{IEEEtran}
\IEEEoverridecommandlockouts

\usepackage{cite}
\usepackage{amsmath,amssymb,amsfonts}
\usepackage{algorithmic}
\usepackage{graphicx}
\usepackage{textcomp}
\usepackage{xcolor}
\usepackage{array}
\usepackage{mdwmath}
\usepackage{mdwtab}
\usepackage{eqparbox}
\usepackage{url}
\usepackage{tabularx}
\usepackage{subcaption}
\usepackage{mathrsfs}
\usepackage{rsfso}
\usepackage{multirow}
\usepackage{amssymb}
\usepackage{cite}
\usepackage{amsmath,amssymb,amsfonts}
\usepackage{algorithmic}
\usepackage{textcomp}
\usepackage{xcolor}
\def\BibTeX{{\rm B\kern-.05em{\sc i\kern-.025em b}\kern-.08em
    T\kern-.1667em\lower.7ex\hbox{E}\kern-.125emX}}
\begin{document}

\title{Unification of Secret Key Generation and Wiretap Channel Transmission\\

\thanks{\copyright IEEE. This work was supported in part by the U.S. Department of Defense under W911NF-20-2-0267. The views and conclusions contained in this
document are those of the authors and should not be interpreted as representing the official policies, either
expressed or implied, of  the U.S. Government. The U.S. Government is
authorized to reproduce and distribute reprints for Government purposes notwithstanding any copyright
notation herein.}
}

\author{\IEEEauthorblockN{Yingbo Hua and Md Saydur Rahman}
\IEEEauthorblockA{\textit{Department of Electrical and Computer Engineering} \\
\textit{University of California at Riverside}\\
Riverside, CA, 92521, USA \\
yhua@ucr.edu and mrahm054@ucr.edu
}}

\maketitle
\begin{abstract}
This paper presents further insights into a recently developed round-trip communication scheme called ``Secret-message Transmission by Echoing Encrypted Probes (STEEP)''. A legitimate wireless channel between a multi-antenna user (Alice)  and a single-antenna user (Bob) in the presence of a multi-antenna eavesdropper (Eve) is focused on.  STEEP does not require full-duplex, channel reciprocity or Eve's channel state information, but is able to yield a positive secrecy rate in bits per channel use between Alice and Bob in every channel coherence period as long as Eve's receive channel is not noiseless. This secrecy rate does not diminish as coherence time increases. Various statistical behaviors of STEEP's secrecy capacity due to random channel fading are also illustrated.
\end{abstract}


\section{Introduction}

Establishing a secret key between two (or more) nodes in a network is crucial for wide ranges of security applications, including authenticity, confidentiality and integrity for subsequent communications between the nodes. Secret keys are also important for security in artificial intelligence and machine learning. A crucial key-generation tool heavily relied upon by our Internet-based society is known as public key infrastructure (PKI), which is based on pair of public and private (asymmetric) keys generated by each node, and which is known to be not information-theoretically (IT) secure as defined in \cite{Shannon1948}. It remains possible that advanced computing algorithms and/or devices developed in the future could be capable of destroying PKI. It is therefore important for us as researchers to develop IT-secure methods for key generations and distributions.

To be able to transmit a secret (including secret key, secret message or secret information) from Alice to Bob in the presence of an eavesdropper (Eve) in any channel coherence period, the classic wiretap channel theory in \cite{Wyner1975} and \cite{Csiszar1978} requires that the main channel (i.e., from Alice to Bob) is stronger than Eve's channel (i.e., from Alice to Eve). For wireless fading channels, one can take the advantage of the situations where the main channel may become stronger than Eve's channel in some random time intervals \cite{Feng2022}. However, to the authors' knowledge, if the main channel is never stronger than Eve's channel in any coherence period, then none of the prior schemes based on wiretap channel model yields a positive secrecy rate.

To be able to generate  a secret key between Alice and Bob regardless of the strength of Eve's channel, many efforts have been made by researchers to exploit the reciprocal nature of (some) wireless channels \cite{Wilson2007} and \cite{Wallace2010}. But the secret-key capacity based on reciprocal channels is limited by channel coherence time. In other words, the secret-key capacity is inversely proportional to the channel coherence time. In most practical situations such as applications in Internet-of-Things where the coherence time is relatively long, the secret-key capacity based on reciprocal channels is very limited.

The limitation from reciprocal channels has driven researchers to explore alternative approaches.  For example, the authors in \cite{HessamMahdavifar2020} and \cite{Li2022} proposed the use of random pilots of a long length to probe a reciprocal channel with the hope that their approaches would yield a secret-key rate not limited by the channel coherence time. Those  works motivated the recent contributions shown in \cite{Hua2023} and \cite{HuaMaksud2023March} where the lower and upper bounds on secret-key capacity established by Maurer, Ahlswede and Csiszar (MAC) in \cite{Maurer1993} and \cite{Ahlswede1993} are applied to the data sets collected from MIMO channels driven by random probes. It is shown in  \cite{Hua2023} that the secure degree of freedom (i.e., degree of freedom of secret-key capacity) of the approaches in \cite{HessamMahdavifar2020} and \cite{Li2022} is no different from the prior approaches based on reciprocal channel responses. Furthermore, the work in \cite{Hua2023} discovered that if and only if Alice or Bob has more antennas than Eve, the secure degree of freedom (in bits per probing sample interval per doubling of power) is positive.

According to \cite{Hua2023}, if Alice and Bob each have a single antenna, then their secure degree of freedom against Eve (with one or more antennas) relative to transmitted power is zero. This result however does not provide the complete picture of the secret-key capacity achievable from a SISO channel against Eve. More recently, it is shown in \cite{Hua2023b} that the secret-key capacity (in bits per probing sample) based on the data sets from a SISO (non-reciprocal) channel driven by random probes is
\begin{equation}\label{eq:C_key}
  C_{key} = \mathbb{E}\left \{\log\left ( 1+\frac{\texttt{SNR}_m}{1+\texttt{SNR}_e}\right )\right\}
\end{equation}
where $\texttt{SNR}_m$ is the signal-to-noise ratio (SNR) at the receiver of the SISO main channel during channel probing, and $\texttt{SNR}_e$ is the corresponding SNR at (multi-antenna) Eve. This capacity is always positive provided that  $\texttt{SNR}_m>0$ and $\texttt{SNR}_e<\infty$.

Furthermore, it is shown in \cite{Hua2023b} that a subsequent transmission scheme assisted by the data sets collected from channel probing allows the application of any established wiretap channel transmission scheme to achieve the secrecy capacity shown in \eqref{eq:C_key}. This new scheme is called ``secret-message transmission by echoing encrypted probes (STEEP)'', which is an important unification of the prior principles and theories of wiretap channel transmission and secret key generation.

In this paper, we show further insights into STEEP with focus on a wireless network consisting of multi-antenna Alice, single-antenna Bob and multi-antenna Eve. In addition to analytical insights, this paper also shows useful observations from computer simulations. It is further demonstrated that STEEP is capable to yield a positive secrecy rate within each channel coherence period even if Eve's receive channel is stronger at all times than the main/legitimate channel. This property of STEEP is not available from any prior wiretap-channel schemes or reciprocal-channel based schemes. The impact of random channel fading on the performance of STEEP is also shown with comparison to a conventional half-duplex two-way scheme subject to the same power allocations.

\section{The Principle of STEEP}

The principle of STEEP \cite{Hua2023b} consists of two phases of interdependent operations: phases 1 and 2.

In phase 1, a node (Alice) sends random probing symbols (also called probes) over a probing channel to another node (Bob). In this phase, Bob obtains some estimates of the probes, which could be noisy. While the estimates of the probes by Eve cannot be noiseless, they are allowed to be less noisy than those by Bob.

In phase 2, Bob echoes back his estimated probes encrypted by a secret message meant for Alice via a return channel. Since Alice knows the exact probes, the \emph{effective} wiretap channel system from Bob to Alice and Eve, relative to the secret message from Bob, is such that the \emph{effective} return channel from Bob to Alice is stronger than that from Bob to Eve subject to a sufficient amount of power from Bob.

For a SISO probing channel from Alice to Bob, along with high capacities from Bob to both Alice and Eve, the secrecy capacity of STEEP can be made \cite{Hua2023b} to approach \eqref{eq:C_key}. Next, we consider an application of STEEP to a MISO wireless channel between Alice and Bob.

\section{Application of STEEP to MISO Wireless Channel}

\begin{figure}[h]
    \centering
    \includegraphics[width=5.5cm,height=4cm]{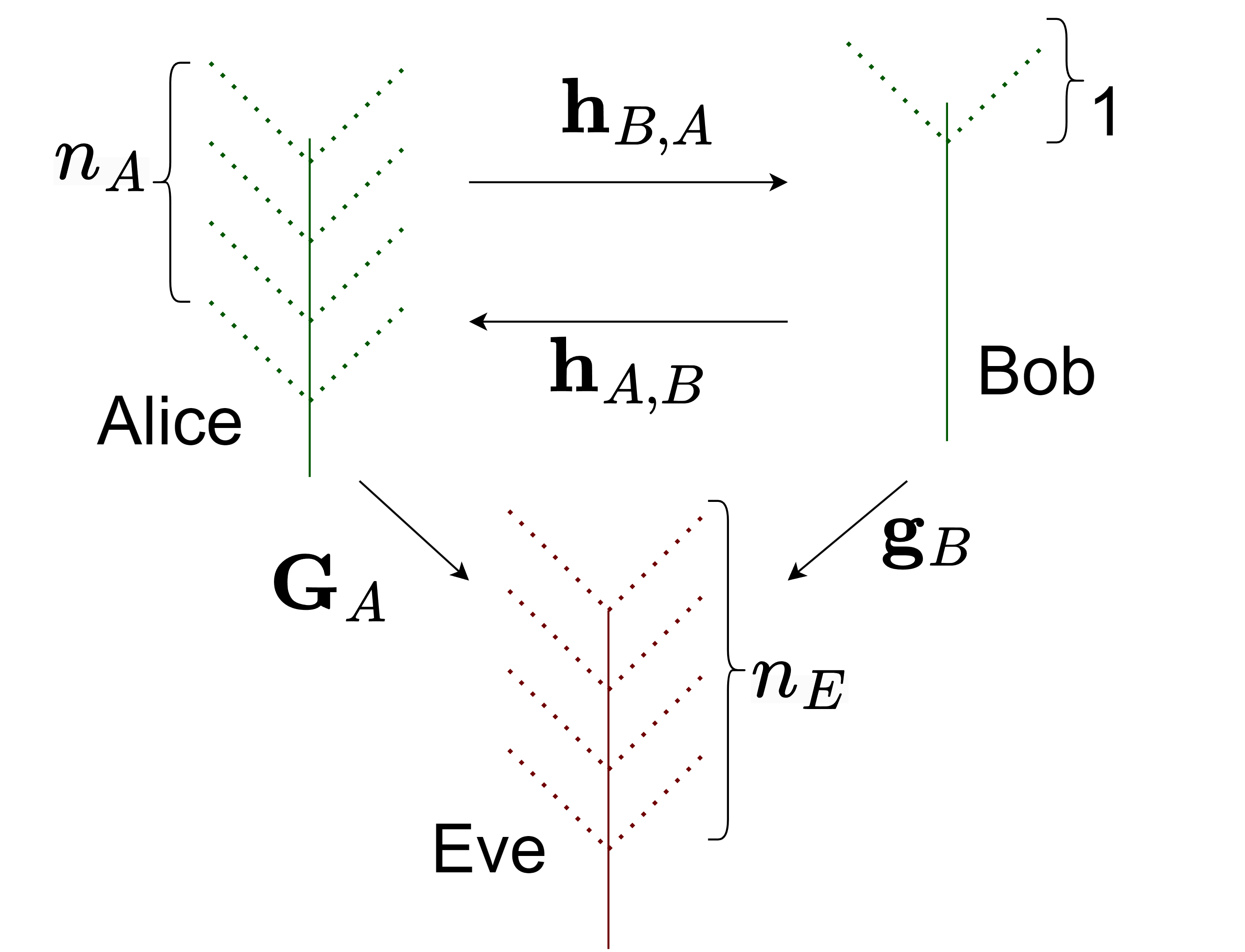}
    \centering
    \caption{A (half-duplex) wireless network with multi-antenna Alice, single-antenna Bob and multi-antenna Eve.}
    \label{fig:mesh2}
\end{figure}

A MISO wireless channel is very common between a base station (or access point) and a mobile node (or user equipment), which is illustrated in Fig. \ref{fig:mesh2}. Here we treat the downlink (from Alice to Bob) as the probing channel, and the uplink (from Bob to Alice) as the return channel. We will assume that the channel responses between Alice and Bob are known to both of them, and all channel responses in the network are known to Eve.

It is important to note that according the theory developed in \cite{Hua2023} the channel probing should always be applied from the node with more antennas to the node with less antennas. In this way, the highest secure degree of freedom is achieved. Specifically, if (independent and identically distributed or i.i.d.) random probing symbols are transmitted from Alice with $n_A$ antennas to Bob with $n_B$ antennas over $m_A\geq n_A$ probing sample intervals, and also another set of random probing symbols are transmitted from Bob to Alice over additional $m_B\geq n_B$ probing sample intervals, then the secure degree of freedom of the secret-key capacity (in bits per probing sample interval) based on the data sets collected from the channel probing is (according to \cite{Hua2023})
\begin{align}\label{eq:SDoF}
  &\texttt{SDoF} = \frac{1}{m_A+m_B}[\min(n_B,(n_A-n_E)^+)(m_A-n_A)\notag\\
  &\,\, +
  \min(n_A,(n_B-n_E)^+)(m_B-n_B)
  +n_An_B\delta]
\end{align}
where $n_E$ is the number of antennas on Eve, $(x)^+=\max(0,x)$, $\delta=1$ if the channel is perfectly reciprocal, and $\delta=0$ if there is no perfectly reciprocal channel response parameter.
The above SDoF does not change if the probes from Alice during $n_A$ probing sample intervals are public and the probes from Bob during $n_B$ probing sample intervals are public.
So, if $m_A+m_B$ is fixed and $n_A>n_B$, then the SDoF is maximized by maximizing $m_A$ and minimizing $m_B$ (i.e., choosing $m_B=n_B$), which is equivalent to one-way random channel probing from Alice to Bob. In the sequel, we use $n_B=1$.

In phase 1 of STEEP, Alice sends random probes, denoted by the matrix $\sqrt{P_A/n_A}\mathbf{X}_A\in \mathbb{C}^{n_A\times m}$ with  $\mathbf{X}_A =[\mathbf{x}_A(1),\cdots ,\mathbf{x}_A(m)]$. Here $m$ is the total number of random probing sample intervals, and all entries in $\mathbf{X}_A$ are i.i.d. (complex circular Gaussian) $CN(0,1)$ random variables. Consequently, the signals received by Bob and Eve in phase 1 can be written as
\begin{equation}\label{}
\mathbf{y}_B^{T} = \sqrt{P_A/n_A}\mathbf{h}_{B,A}^{T}\mathbf{X}_A + \mathbf{w}_B^T \in \mathbb{C}^{1\times m},
    \end{equation}
    \begin{equation}\label{}
    \mathbf{Y}_{E,A} =  \sqrt{P_A/n_A}\mathbf{G}_A\mathbf{X}_A + \mathbf{W}_{E,A}\in\mathbb{C}^{n_E\times m}.
    \end{equation}
Here $\mathbf{h}_{B,A}$ is the channel vector from Alice to Bob, $\mathbf{G}_A$ is the channel matrix from Alice to Eve, and $\mathbf{w}_B^T$ and $\mathbf{W}_{E,A}$ are the noises. We will let the entries in $\mathbf{w}_B^T$ be i.i.d. $CN(0,\sigma_B^2)$ and those in $\mathbf{W}_{E,A}$ be i.i.d. $CN(0,\sigma_{E,A}^2)$.

In phase 2 of STEEP, Bob echoes back the probes encrypted by a secret message for Alice. Here we consider the (not optimized) signal sent by Bob: $\sqrt{P_B'} (\mathbf{s}^{T}+ \sqrt{n_A/P_A}\mathbf{y}_B^T)$ with $\mathbf{s}^T=[s(1),\cdots,s(m)]$ containing the secret message for Alice. We will assume that all entries in $\mathbf{s}^T$ are i.i.d. $CN(0,1)$.

Note that while $P_A$ is a fair representation of the transmission power by Alice in phase 1, $P_B'$ is only a ``reference power'' used by Bob in phase 2. The actual power consumed by Bob is
\begin{equation}\label{}
  P_B=P_B'+P_B'\|\mathbf{h}_{B,A}\|^2+\frac{n_A}{P_A}\sigma_B^2.
\end{equation}
 Since $\mathbf{h}_{B,A}$ is assumed to be public, we can assume that $P_A$, $P_B'$ and $P_B$ are all public.

Then the signals received by Alice and Eve via the return channels in phase 2 can be written as
    \begin{equation}\label{}
    \mathbf{Y}_A =  \sqrt{P_B'}\mathbf{h}_{A,B}(\mathbf{s}^T
     + \sqrt{n_A/P_A}\mathbf{y}_B^T)+ \mathbf{W}_A\in\mathbb{C}^{n_A\times m},
    \end{equation}
    \begin{equation}\label{}
    \mathbf{Y}_{E,B} =  \sqrt{P_B'}\mathbf{g}_B (\mathbf{s}^T + \sqrt{n_A/P_A}\mathbf{y}_B^T) + \mathbf{W}_{E,B}\in\mathbb{C}^{n_E\times m}.
    \end{equation}
Here $\mathbf{h}_{A,B}$ is the  channel vector from Bob to Alice, $\mathbf{g}_B$ is the channel vector from Bob to Eve, and $\mathbf{W}_A$ and $\mathbf{W}_{E,B}$ are the noises. We will let the entries of $\mathbf{W}_A$ be i.i.d. $CN(0,\sigma_A^2)$ and those of $\mathbf{W}_{E,B}$ be i.i.d. $CN(0,\sigma_{E,B}^2)$ .

\subsection{Analysis of the signals received by Alice}
Since Alice knows $\sqrt{n_{A}P_{B}'/P_{A}}\mathbf{h}_{A,B}\mathbf{h}_{B,A}^{T}\mathbf{X}_{A}$, subtracting it from $\mathbf{Y}_A$ yields
\begin{equation}
    \mathbf{Y}_A'=\sqrt{P_B'}\mathbf{h}_{A,B}\mathbf{s}^T + \mathbf{W}_A'
\end{equation}
with
$\mathbf{W}_A'=\sqrt{n_AP_B'/P_A}\mathbf{h}_{A,B}\mathbf{w}_B^{T}+ \mathbf{W}_A$.
The $k$th column of $\mathbf{Y}_A'$ is written as
\begin{equation}
    \mathbf{y}_A'(k)=\sqrt{P_B'}\mathbf{h}_{A,B}s(k) + \mathbf{w}_A'(k),
\end{equation}
and the sufficient statistic from  $\mathbf{y}_A'(k)$ for $s(k)$ is
\begin{align}\label{eq:main_return}
  &r_A(k) \doteq \frac{1}{\sqrt{P_B'}\|\mathbf{h}_{A,B}\|^2}\mathbf{h}_{A,B}^H\mathbf{y}_A'(k)\notag\\
  &=s(k)+v_A(k)
\end{align}
with $v_A(k) = \sqrt{n_A/P_A}w_B(k)+\frac{\mathbf{h}_{A,B}^H\mathbf{w}_A(k)}{\sqrt{P_B'}\|\mathbf{h}_{A,B}\|^2}$ which, conditioned on $\mathbf{h}_{A,B}$, is $CN(0,\sigma_{v,A}^2)$ with
\begin{equation}\label{}
  \sigma_{v,A}^2 = \frac{n_A}{P_A}\sigma_B^2+\frac{\sigma_A^2}{P_B'\|\mathbf{h}_{A,B}\|^2}.
\end{equation}

\subsection{Analysis of the signals received by Eve}
Note that conditioned on the channel responses, the columns of $\mathbf{Y}_{E,A}$ are independent of each other. The same is true for $\mathbf{Y}_{E,B}$.
Let the $k$th column of $\mathbf{Y}_{E,A}$ and that of $\mathbf{Y}_{E,B}$ be written as
 \begin{equation}\label{}
    \mathbf{y}_{E,A}(k) =  \sqrt{P_A/n_A}\mathbf{G}_A\mathbf{x}_A(k) + \mathbf{w}_{E,A}(k),
    \end{equation}
      \begin{align}
    &\mathbf{y}_{E,B}(k) \notag\\
    &=  P_B'^{1/2}\mathbf{g}_B (s(k) + (n_A/P_A)^{1/2}y_B(k)) + \mathbf{w}_{E,B}(k)
    \notag\\
    &=P_B'^{1/2}\mathbf{g}_B s(k) +P_B'^{1/2}\mathbf{g}_B\mathbf{h}_{B,A}^T\mathbf{\hat x}_A(k)+\mathbf{w}_{E,B}'(k),
    \end{align}
    with
\begin{align}
&\mathbf{w}_{E,B}'(k)=\sqrt{P_B'}\mathbf{g}_B\mathbf{h}_{B,A}^T\Delta \mathbf{x}_A(k)\notag\\
&\,\,+\sqrt{n_AP_B'/P_A}\mathbf{g}_Bw_B(k)+\mathbf{w}_{E,B}(k).
\end{align}
 Here $\mathbf{\hat x}_A(k)+\Delta \mathbf{x}_A(k)=\mathbf{x}_A(k)$, and $\mathbf{\hat x}_A(k)$ is the minimum mean squared error (MMSE) estimate of $\mathbf{x}_A(k)$ from $\mathbf{y}_{E,A}(k)$. It follows that
\begin{align}\label{}
  &\mathbf{R}_{\Delta x}\doteq\mathbb{E}\{\Delta \mathbf{x}_A(k)\Delta \mathbf{x}_A(k)^H\}
  \notag\\
  &=\mathbf{I}_{n_A}-\frac{P_A}{n_A}\mathbf{G}_A^H\left (\frac{P_A}{n_A}\mathbf{G}_A\mathbf{G}_A^H
  +\sigma_{E,A}^2\mathbf{I}_{n_E}\right )^{-1}\mathbf{G}_A\notag\\
  &=\left (\frac{P_A}{n_A\sigma_{E,A}^2}\mathbf{G}_A^H\mathbf{G}_A+\mathbf{I}_{n_A} \right )^{-1}.
\end{align}

Then the sufficient statistic from $\mathbf{\hat x}_A(k)$ and $\mathbf{y}_{E,B}(k)$ for $s(k)$ is (a more rigorous treatment is in an upcoming paper):
\begin{align}\label{eq:eve_return}
&r_E(k) \doteq \frac{1}{\sqrt{P_B'}\|\mathbf{g}_B\|^2}\mathbf{g}_B^H\left (\mathbf{y}_{E,B}(k)-\sqrt{P_B'}\mathbf{g}_B\mathbf{h}_{B,A}^T\mathbf{\hat x}_A(k)\right )\notag\\
&=s(k) + v_E(k)
\end{align}
with
\begin{align}
&v_E(k) = \mathbf{h}_{B,A}^T\Delta\mathbf{x}_A(k) +\sqrt{\frac{n_A}{P_A}}w_B(k) \notag\\
&\,\,+\frac{\mathbf{g}_B^H\mathbf{w}_{E,B}(k)}{\sqrt{P_B'}\|\mathbf{g}_B\|^2}
\end{align}
which (conditioned on channel responses) is $CN(0,\sigma_{v,E}^2)$ with
\begin{equation}\label{}
  \sigma_{v,E}^2
  =\mathbf{h}_{B,A}^T\mathbf{R}_{\Delta x}\mathbf{h}_{B,A}^*+\frac{n_A}{P_A}\sigma_B^2+\frac{\sigma_{E,B}^2}{P_B'\|\mathbf{g}_B\|^2}.
\end{equation}

\subsection{Analysis of the secrecy rate of the effective return channels}
The above described STEEP forms effective return channels from Bob to Alice and Eve, i.e., \eqref{eq:main_return} and  \eqref{eq:eve_return}. Hence the achievable secrecy capacity based on the effective wiretap channel model (with a long coherence time) is known to be $\bar C_{\texttt{STEEP}}\doteq(C_{\texttt{STEEP}})^+$ with
\begin{equation}\label{eq:C_STEEP1}
  C_{\texttt{STEEP}}=\log\left (1+\frac{1}{\sigma_{v,A}^2}\right )-\log\left (1+\frac{1}{\sigma_{v,E}^2}\right ).
\end{equation}
Strictly speaking, this secrecy capacity holds if and only if Alice and Bob  know both $\sigma_{v,A}^2$ and $\sigma_{v,E}^2$. If $\sigma_{v,E}^2$ is unknown to Alice and Bob, a meaningful measure of secrecy is an outage probability for a given secrecy rate $R_s>0$, i.e.,
\begin{align}\label{eq:out_STEEP}
  &O_{\texttt{STEEP}}(R_s)\doteq\texttt{Prob}(\bar C_{\texttt{STEEP}}\leq R_s)\notag\\
  &=\texttt{Prob}(C_{\texttt{STEEP}}\leq R_s).
\end{align}

\subsubsection{The case of large $P_B$}\label{sec:large_PB}
If Bob applies a large power $P_B$ (or $P_B'$) such that $\frac{n_A}{P_A}\sigma_B^2\gg \frac{\sigma_A^2}{P_B'\|\mathbf{h}_{A,B}\|^2}$ and $\frac{n_A}{P_A}\sigma_B^2\gg\frac{\sigma_{E,B}^2}{P_B'\|\mathbf{g}_B\|^2}$ (or equivalently if
$\sigma_{v,A}^2 \approx \frac{n_A}{P_A}\sigma_B^2$ and $\sigma_{v,E}^2
  \approx\mathbf{h}_{B,A}^T\mathbf{R}_{\Delta x}\mathbf{h}_{B,A}^*+\frac{n_A}{P_A}\sigma_B^2$), then
  \begin{align}\label{eq:C_STEEP2}
  &C_{\texttt{STEEP}}\approx\log\left (1+\frac{P_A}{n_A\sigma_B^2}\right )\notag\\
  &\,\,-\log\left (1+\frac{P_A}{n_A\sigma_B^2}\cdot\frac{1}{1+\frac{P_A}{n_A\sigma_B^2}\mathbf{h}_{B,A}^T\mathbf{R}_{\Delta x}\mathbf{h}_{B,A}^*}\right )\notag\\
  &=\log\left (1+\frac{\alpha}{1+\frac{1+\frac{1}{\alpha}}{\beta}}\right )>0
\end{align}
with $\alpha=\frac{P_A}{n_A\sigma_B^2}$ and
\begin{equation}\label{eq:beta}
  \beta=\mathbf{h}_{B,A}^T\mathbf{R}_{\Delta x}\mathbf{h}_{B,A}^*.
\end{equation}
A larger $P_B$ relative to $P_A$ means a higher quality channel from Bob to Alice than from Alice to Bob. In this case, we see that the secrecy capacity $C_{\texttt{STEEP}}$ stays positive as long as $\beta>0$. Note that for a sufficiently large $P_B$, \eqref{eq:C_STEEP2} holds for both static and block-fading channels even if Eve's channels from Alice and Bob are stronger (due to $n_E>n_A$ for example) than the main channels between Alice and Bob.

\subsubsection{The case of $n_A>n_E$}
In this case, the eigenvalue decomposition of $\mathbf{G}_A^H\mathbf{G}_A$ is $\mathbf{Q}\boldsymbol{\Lambda}\mathbf{Q}^H$ with $\mathbf{Q}$ being a $n_A\times n_A$ unitary matrix and $\boldsymbol{\Lambda}=diag(\lambda_1,\cdots,\lambda_{n_E},0,\cdots,0)$. Then
\begin{equation}\label{}
  \beta = \mathbf{h}_{B,A}^T\mathbf{Q}(\alpha'\boldsymbol{\Lambda}+
  \mathbf{I}_{n_A})^{-1}\mathbf{Q}^H\mathbf{h}_{B,A}^*
\end{equation}
with $\alpha'=\frac{P_A}{n_A\sigma_{E,A}^2}$. For a large $P_A$,
\begin{equation}\label{}
  \beta \approx \|\mathbf{h}_{B,A}^T\mathbf{Q}_1\|^2
\end{equation}
where $\mathbf{Q}_1$ consists of the last $n_A-n_E$ columns of $\mathbf{Q}$, and hence  $\beta$ is invariant to large $P_A$. In this case, the degree of freedom of $C_{\texttt{STEEP}}$ in \eqref{eq:C_STEEP2} relative to $\log P_A$ equals one, i.e., $\lim_{P_A\to\infty}\frac{C_{\texttt{STEEP}}}{\log P_A}=1$. (Note that $\alpha$ is proportional to $P_A$.) This is consistent with \eqref{eq:SDoF} with $m_A>n_A>n_E$ and $m_B=n_B=1$.

Note that if the channel probing was conducted from Bob (with one antenna) to Alice (with $n_A\geq 1$ antenna), the corresponding contribution of degree of freedom of secrecy (against Eve with $n_E\geq 1$ antennas) would be zero.

\subsubsection{The case of $n_A\leq n_E$}
In this case, $\boldsymbol{\Lambda}=diag(\lambda_1,\cdots,\lambda_{n_E})>0$. For
a large $P_A$, we have
$\beta\approx 0$
but
\begin{align}\label{}
  &\alpha\beta \approx \frac{\sigma_{E,A}^2}{\sigma_B^2}\mathbf{h}_{B,A}^T\mathbf{Q}\boldsymbol{\Lambda}^{-1}
  \mathbf{Q}^H\mathbf{h}_{B,A}^*\notag\\
  &=\frac{\sigma_{E,A}^2}{\sigma_B^2}\mathbf{h}_{B,A}^T
  (\mathbf{G}_A^H\mathbf{G}_A)^{-1}\mathbf{h}_{B,A}^*>0.
\end{align}
Namely, $\alpha\beta$ stays positive and invariant to large $P_A$ while $\beta$ and $1/\alpha$ converge to zero as $P_A$ increases. In this case, for large $P_A$, $C_{\texttt{STEEP}}$ in \eqref{eq:C_STEEP2} becomes
 \begin{align}\label{eq:C_STEEP3}
  &C_{\texttt{STEEP}}\approx\log\left (1+\frac{\sigma_{E,A}^2}{\sigma_B^2}\mathbf{h}_{B,A}^T
  (\mathbf{G}_A^H\mathbf{G}_A)^{-1}\mathbf{h}_{B,A}^*\right )>0
\end{align}
which is invariant to $P_A$. In this case of $n_A\leq n_E$, the degree of freedom of $C_{\texttt{STEEP}}$ relative to $\log P_A$ is zero, which is expected according to \eqref{eq:SDoF}.

\subsubsection{The case of arbitrary $P_B$}

In this case, in order for $C_{\texttt{STEEP}}$ in \eqref{eq:C_STEEP1} to be non-positive, we must have $\sigma_{v,A}^2\geq \sigma_{v,E}^2$ or equivalently
\begin{equation}\label{eq:outage}
  \frac{\sigma_A^2}{\|\mathbf{h}_{A,B}\|^2}- \frac{\sigma_{E,B}^2}{\|\mathbf{g}_B\|^2}\geq P_B'\beta.
\end{equation}
We will refer to this condition as the natural outage. It is clear that $A\doteq\frac{\sigma_A^2}{\|\mathbf{h}_{A,B}\|^2}$ is the (normalized) return channel attenuation for users (from Bob to Alice) while $E\doteq\frac{\sigma_{E,B}^2}{\|\mathbf{g}_B\|^2}$ is the (normalized) return channel attenuation for Eve (from Bob to Eve). If $A$ is no larger than $E$, the natural outage does not happen. We also see that for any given $A$ and $E$, there is a finite threshold for $P_B'$ beyond which the natural outage does not happen.

Since $\mathbf{g}_B$ is unknown to users, so is $E$ in general. Since $\mathbf{G}_A$ is unknown to users, so is $\beta$ in general. Therefore, the natural outage for any given $P_A$ and $P_B$  is in general a random event. The probability of the natural outage along with other more general properties will be discussed in section \ref{sec:simulation}.

\section{A Conventional Half-Duplex Two-Way Scheme}

For comparison purpose, let us now consider a conventional half-duplex two-way scheme where Alice sends a secret to Bob in phase 1, and Bob sends another secret to Alice in phase 2. We will use the same power consumptions by Alice and Bob and the same channel parameters as used for STEEP.

In phase 1, the optimal waveform to be transmitted is known to be $\sqrt{P_A}\frac{\mathbf{h}_{B,A}^*}{\|\mathbf{h}_{B,A}\|}s_A(k)$ with $s_A(k)$ being i.i.d. $CN(0,1)$. Then the SNRs at Bob and Eve are
\begin{equation}\label{}
  \texttt{SNR}_B=\frac{P_A\|\mathbf{h}_{B,A}\|^2}{\sigma_B^2},
\end{equation}
\begin{equation}\label{}
  \texttt{SNR}_{E,A}=\frac{P_A\|\mathbf{g}_A\|^2}{\sigma_{E,A}^2}
\end{equation}
with $\mathbf{g}_A=\mathbf{G}_A\frac{\mathbf{h}_{B,A}^*}{\|\mathbf{h}_{B,A}\|}$.
The secrecy capacity in phase 1 is  $\bar C_1=(C_1)^+$ with
\begin{equation}\label{}
  C_1=\log(1+\texttt{SNR}_B)-\log(1+\texttt{SNR}_{E,A}).
\end{equation}
It is obvious that $\bar C_1=0$ for any $P_A$ if $\frac{\|\mathbf{g}_A\|^2}{\sigma_{E,A}^2}\geq \frac{\|\mathbf{h}_{B,A}\|^2}{\sigma_B^2}$.

In phase 2, the optimal waveform to be transmitted by Bob is $\sqrt{P_B}s_B(k)$ with $s_B(k)$ being i.i.d. $CN(0,1)$ (and also independent of $s_A(k)$). The signal received by Alice is $\sqrt{P_B}\mathbf{h}_{A,B}s_B(k)+\mathbf{w}_A$, from which a sufficient statistic is obtained by multiplying it from the left by $\frac{\mathbf{h}_{A,B}^H}{\|\mathbf{h}_{A,B}\|}$. The SNR of the resulting signal is
\begin{equation}\label{}
  \texttt{SNR}_A = \frac{P_B\|\mathbf{h}_{A,B}\|^2}{\sigma_A^2}.
\end{equation}
Similarly, the signal received by Eve is $\sqrt{P_B}\mathbf{g}_Bs_B(k)+\mathbf{w}_{E,B}$, and the SNR of a corresponding sufficient statistic of this signal is
\begin{equation}\label{}
  \texttt{SNR}_{E,B} = \frac{P_B\|\mathbf{g}_B\|^2}{\sigma_{E,B}^2}.
\end{equation}
The secrecy capacity in phase 2 is  $\bar C_2=(C_2)^+$ with
\begin{equation}\label{}
  C_2=\log(1+\texttt{SNR}_A)-\log(1+\texttt{SNR}_{E,B}).
\end{equation}
Similar to $\bar C_1$, we see that $\bar C_2=0$ for any $P_B$ if $\frac{\|\mathbf{g}_B\|^2}{\sigma_{E,B}^2}\geq \frac{\|\mathbf{h}_{A,B}\|^2}{\sigma_A^2}$.

The sum secrecy capacity of the conventional scheme is
\begin{equation}\label{}
  \bar C_{\texttt{conv}}=\bar C_1+\bar C_2.
\end{equation}
Note that $\bar C_{\texttt{conv}}\neq (C_1+C_2)^+$. In fact, $\bar C_{\texttt{conv}}\geq (C_1+C_2)^+$.
We will next compare $\bar C_{\texttt{conv}}$ with $\bar C_{\texttt{STEEP}}$ under random realizations of channel responses. More specifically, we will compute the distribution of the improvement or gain of the secrecy capacity from the conventional to STEEP:
\begin{equation}\label{}
  G_s \doteq \bar C_{\texttt{STEEP}} -\bar C_{\texttt{conv}}.
\end{equation}
Note that both $\bar C_{\texttt{conv}}$ and $\bar C_{\texttt{STEEP}}$ measure the achievable number of secret bits established between Alice and Bob per each sample interval for transmission from Alice to Bob and another sample interval for transmission from Bob to Alice.

Since the SNRs at Eve are generally unknown to Alice and Bob, we will also compare $O_{\texttt{STEEP}}(R_s)$ defined in \eqref{eq:out_STEEP} with the following outage probability for the conventional scheme:
\begin{equation}\label{}
  O_{\texttt{conv}}(R_s)\doteq \texttt{Prob}(\bar C_{\texttt{conv}}\leq R_s).
\end{equation}
It is useful to note that $O_{\texttt{conv}}(R_s)\neq \texttt{Prob}(C_1+C_2\leq R_s)$. In fact, $O_{\texttt{conv}}(R_s)\leq \texttt{Prob}(C_1+C_2\leq R_s)$.

\section{Simulation Results}\label{sec:simulation}

In this section, we use simulation to illustrate the secrecy capacity of STEEP shown in \eqref{eq:C_STEEP1}  for some given values of $P_A$ and $P_B$ and subject to random $\mathbf{h}_{B,A}$, $\mathbf{h}_{A,B}=\gamma\mathbf{h}_{B,A}+(1-\gamma)\mathbf{w}$, $\mathbf{g}_B$ and $\mathbf{G}_A$ where $0<\gamma<1$. More specifically, we let all entries of $\mathbf{h}_{B,A}$, $\mathbf{w}$, $\mathbf{g}_B$ and $\mathbf{G}_A$ follow i.i.d. $CN(0,1)$. We also assume $\sigma_B^2=\sigma_A^2=\sigma_{E,A}^2=\sigma_{E,B}^2=1$.

With the above assumptions of the channel responses, $\|\mathbf{h}_{A,B}\|^2$ and $\|\mathbf{g}_B\|^2$ in \eqref{eq:outage} are independent and Chi-square distributed with degrees $2n_A$ and $2n_E$ respectively. But the distribution of $\beta$ in \eqref{eq:outage} or equivalently in \eqref{eq:beta} is more complicated. Also for $0<\gamma<1$, there is a correlation between $\mathbf{h}_{A,B}$ and $\mathbf{h}_{B,A}$, and $\beta$ is also correlated with $\|\mathbf{h}_{A,B}\|^2$. The statistical analysis of $C_{\texttt{STEEP}}$ in \eqref{eq:C_STEEP1} for any given $P_A$ and $P_B$ under the above conditions remains a challenge. In the following, we will present further insights into $C_{\texttt{STEEP}}$ based on computer simulations. We will use $\gamma=0.2$ and $10^5$ independent realizations of all above stated random parameters.

\begin{figure}[ht]
\begin{minipage}[b]{0.45\linewidth}
\centering
\includegraphics[width=\textwidth]{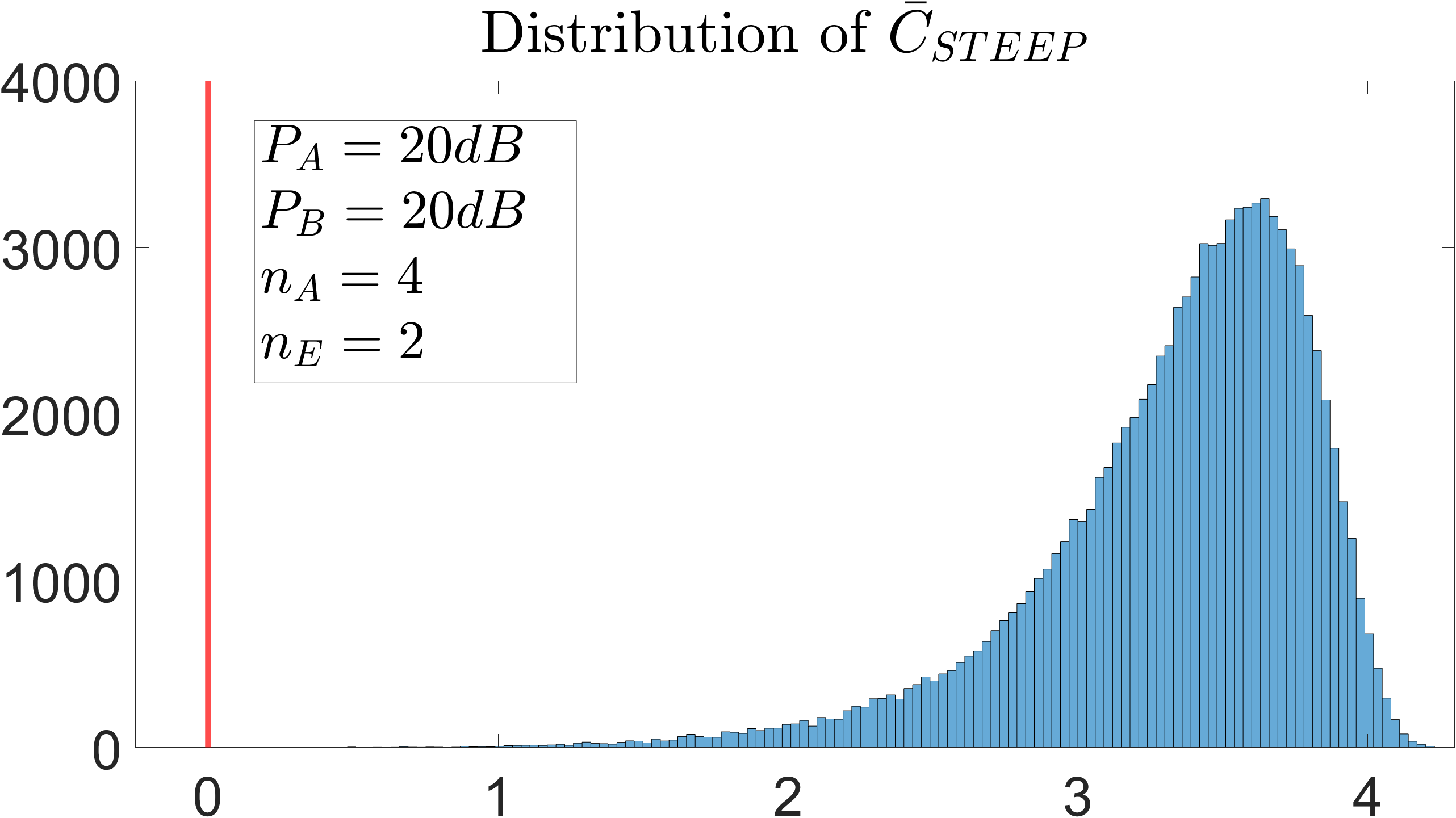}
\subcaption{$n_E=2$ and $P_B=20$dB.}
\label{fig:bar_C_steep_a}
\end{minipage}
\hspace{0.1cm}
\begin{minipage}[b]{0.45\linewidth}
\centering
\includegraphics[width=\textwidth]{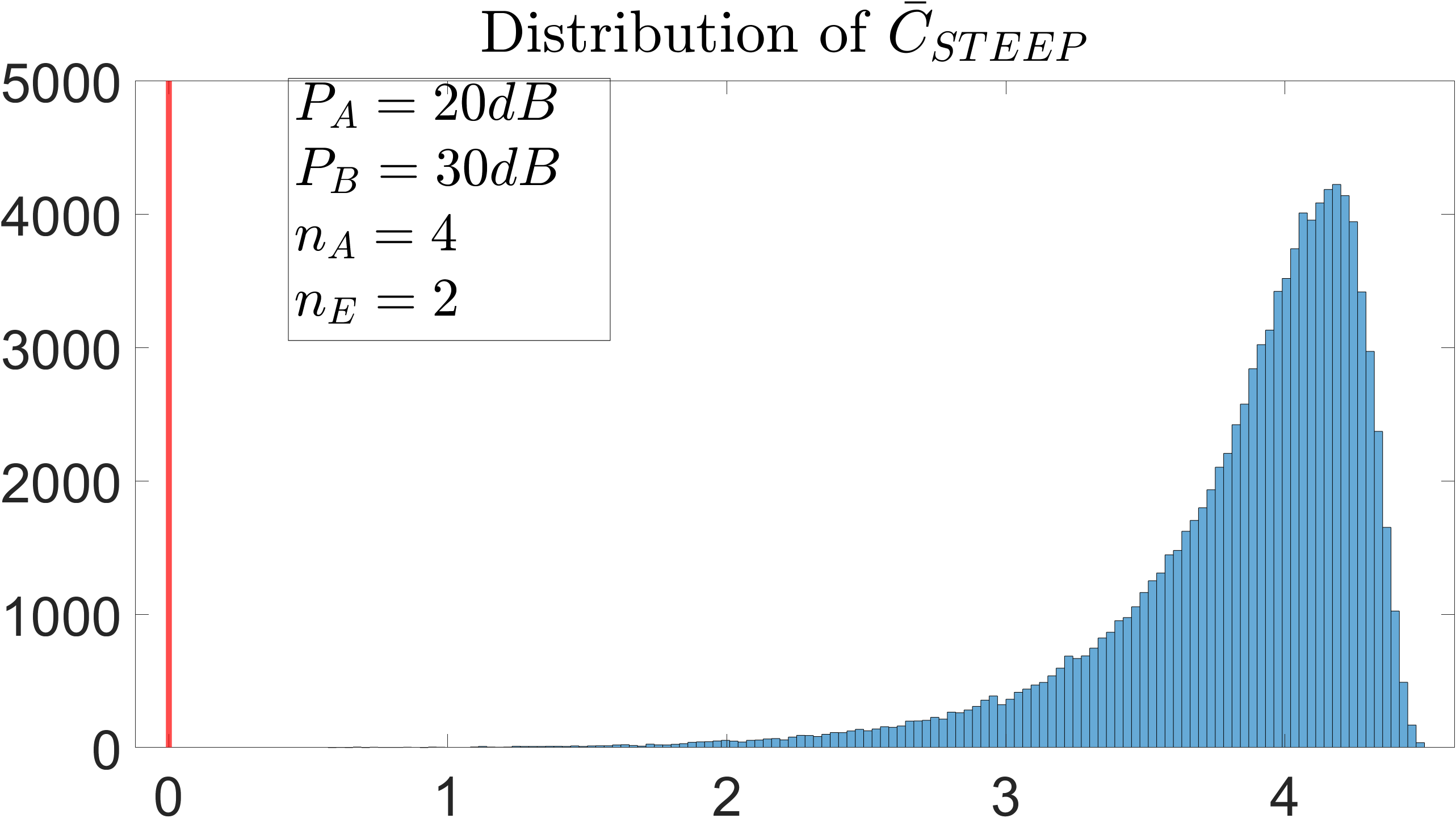}
\subcaption{$n_E=2$ and $P_B=30$dB.}
\label{fig:bar_C_steep_b}
\end{minipage}
\\
\begin{minipage}[b]{0.45\linewidth}
\centering
\includegraphics[width=\textwidth]{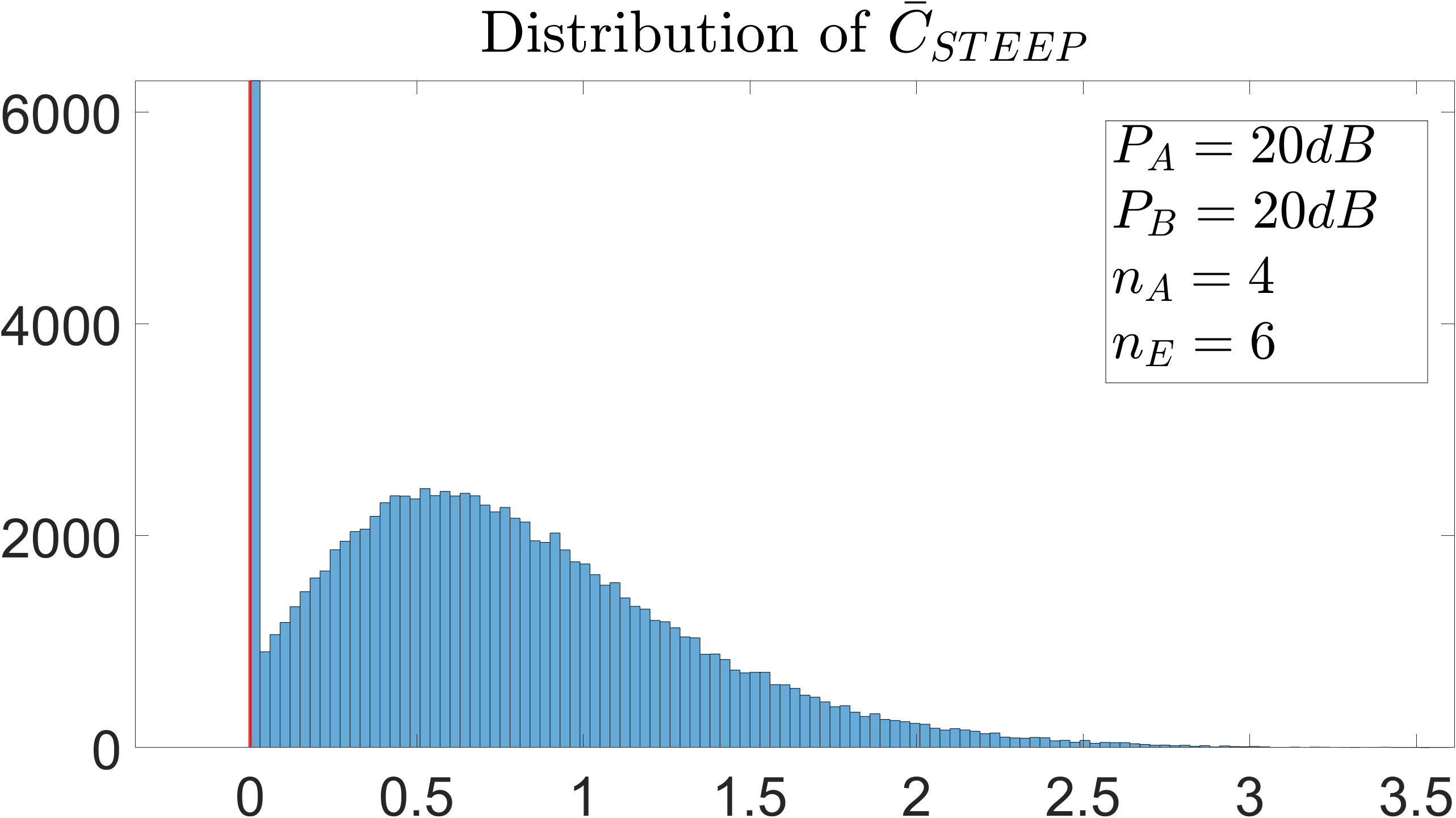}
\subcaption{$n_E=6$ and $P_B=20$dB.}
\label{fig:bar_C_steep_c}
\end{minipage}
\hspace{0.1cm}
\begin{minipage}[b]{0.45\linewidth}
\centering
\includegraphics[width=\textwidth]{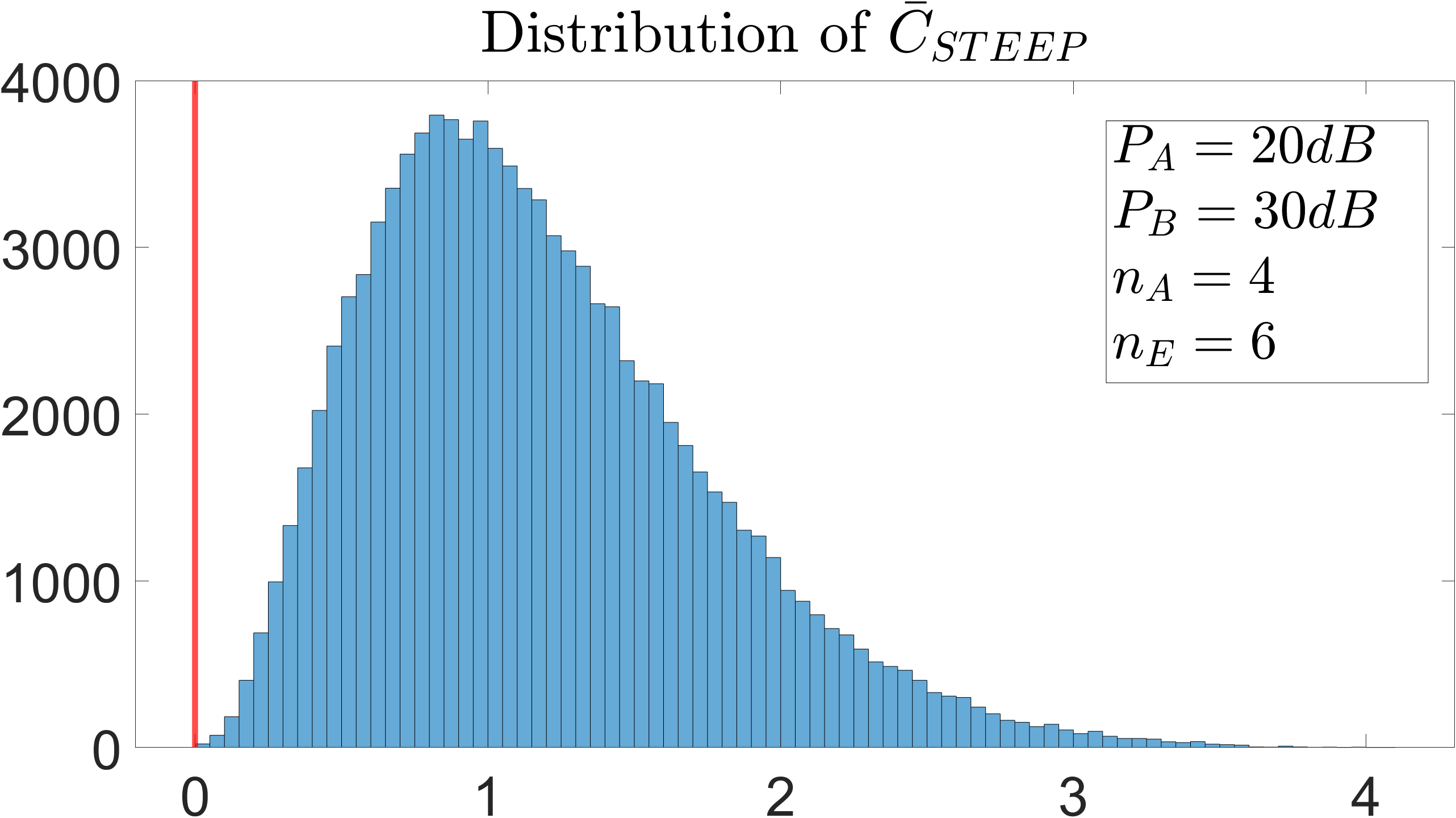}
\subcaption{$n_E=6$ and $P_B=30$dB.}
\label{fig:bar_C_steep_d}
\end{minipage}
\caption{Distribution of $\bar C_{\texttt{STEEP}}$ for $n_A=4$ and $P_A=20$dB.}
\label{fig:bar_C_steep}
\end{figure}

In Fig. \ref{fig:bar_C_steep}, we show the distributions (i.e., histograms) of $\bar C_{\texttt{STEEP}}$ subject to $n_A=4$ and $P_A=20$dB. The upper two plots are for $n_E=2$, and the lower two plots are for $n_E=6$. The left two plots are for $P_B=20$dB, and the right two plots are for $P_B=30$dB. We see that for a larger $P_B$, the probability for the natural outage (i.e., $O_{\texttt{STEEP}}(0)=\texttt{Prob}(\bar C_{\texttt{STEEP}}=0)=\texttt{Prob}( C_{\texttt{STEEP}} \leq 0)$) tends to become much smaller (if not zero), which is justified by the analysis shown in section \ref{sec:large_PB}.
It is also somewhat expected that for a larger $n_E$, the distribution of $\bar C_{\texttt{STEEP}}$ moves to the left. However, it is important to see that for the case of $n_A=4$, $n_E=6$, $P_A=20$dB and $P_B=30$dB (see Fig. \ref{fig:bar_C_steep_d}), the probability of the natural outage for STEEP is still extremely small.

\begin{figure}[ht]
\begin{minipage}[b]{0.45\linewidth}
\centering
\includegraphics[width=\textwidth]{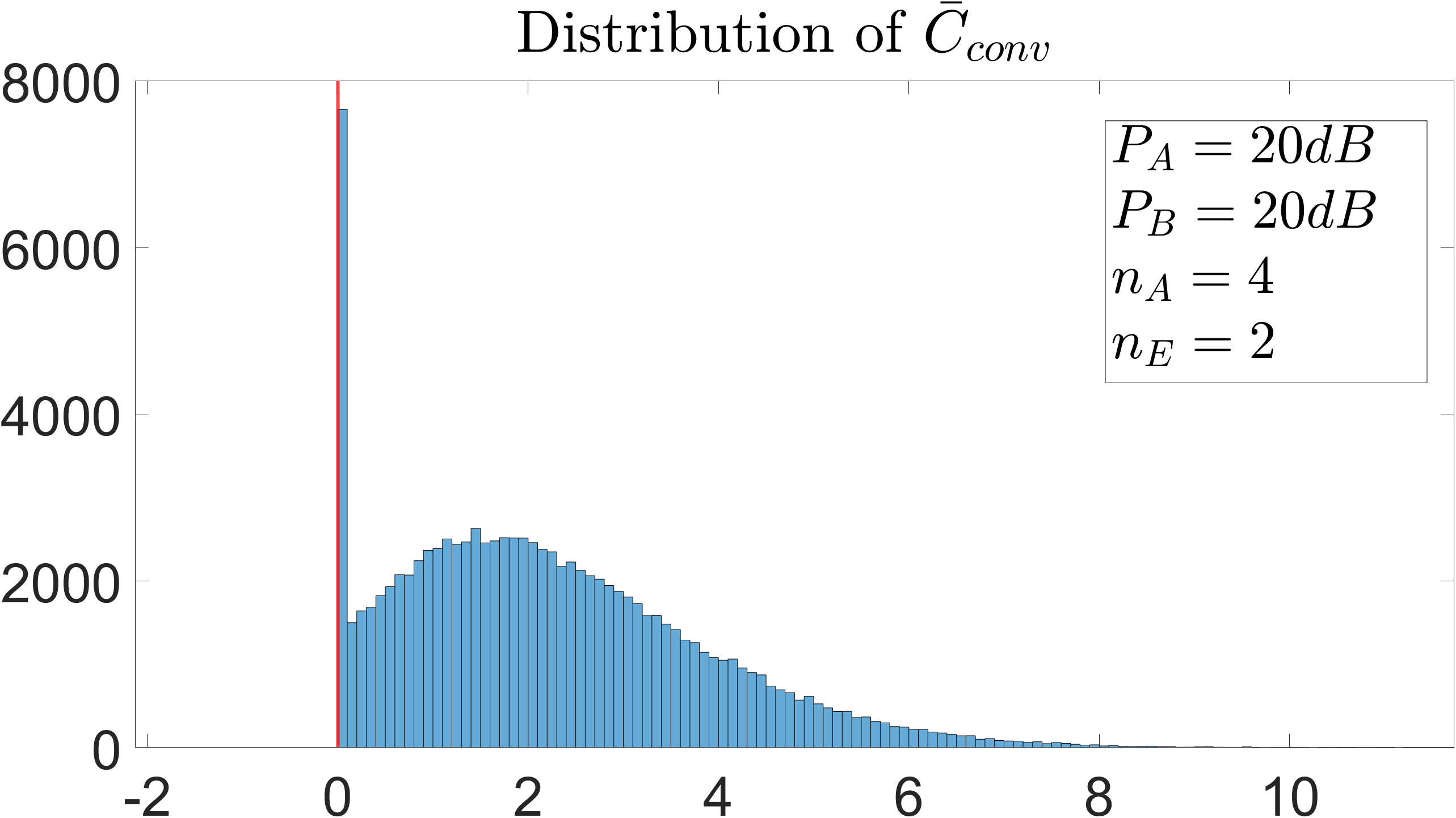}
\subcaption{$n_E=2$ and $P_B=20$dB.}
\label{fig:bar_C_conv_a}
\end{minipage}
\hspace{0.1cm}
\begin{minipage}[b]{0.45\linewidth}
\centering
\includegraphics[width=\textwidth]{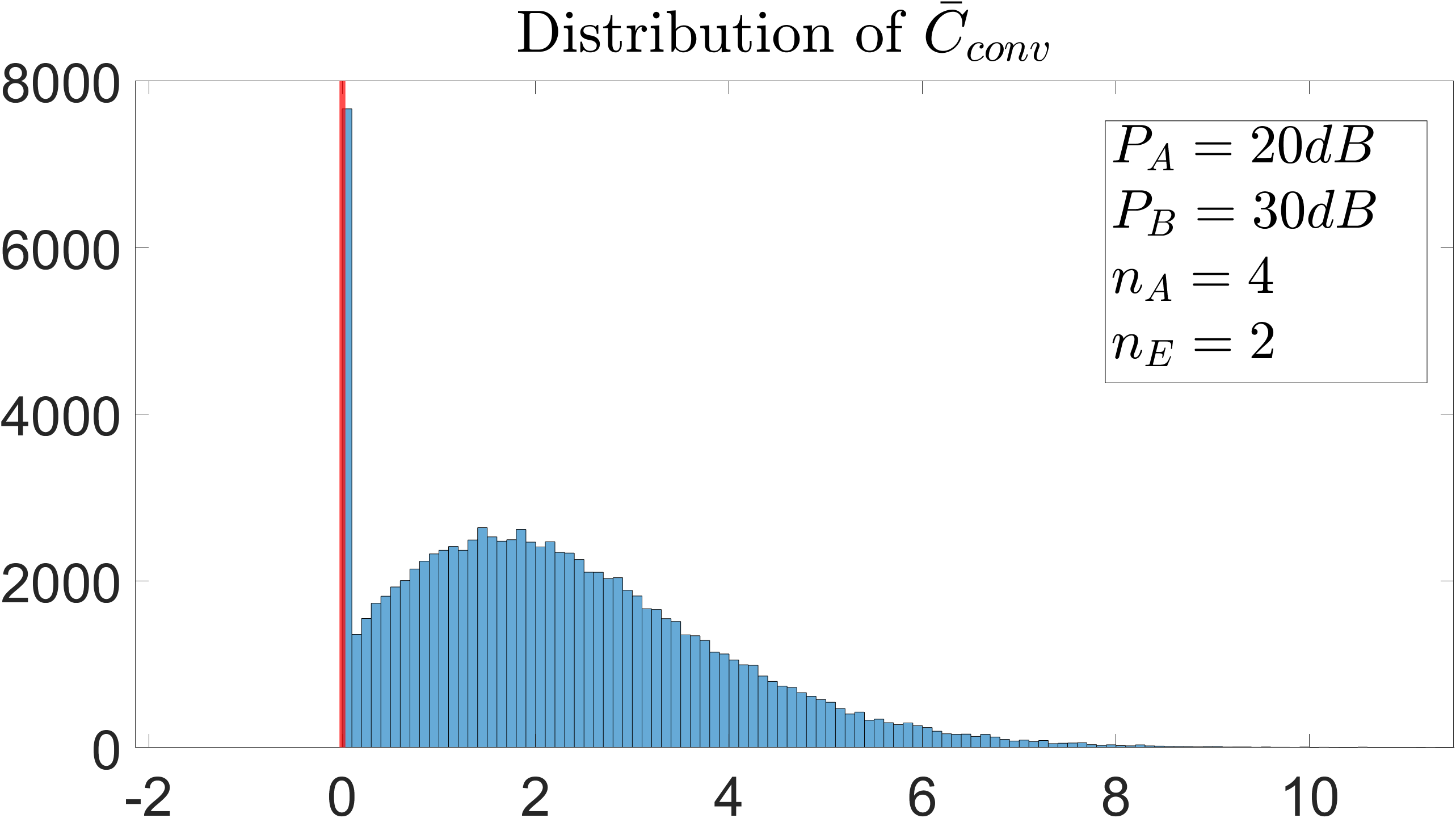}
\subcaption{$n_E=2$ and $P_B=30$dB.}
\label{fig:bar_C_conv_b}
\end{minipage}
\\
\begin{minipage}[b]{0.45\linewidth}
\centering
\includegraphics[width=\textwidth]{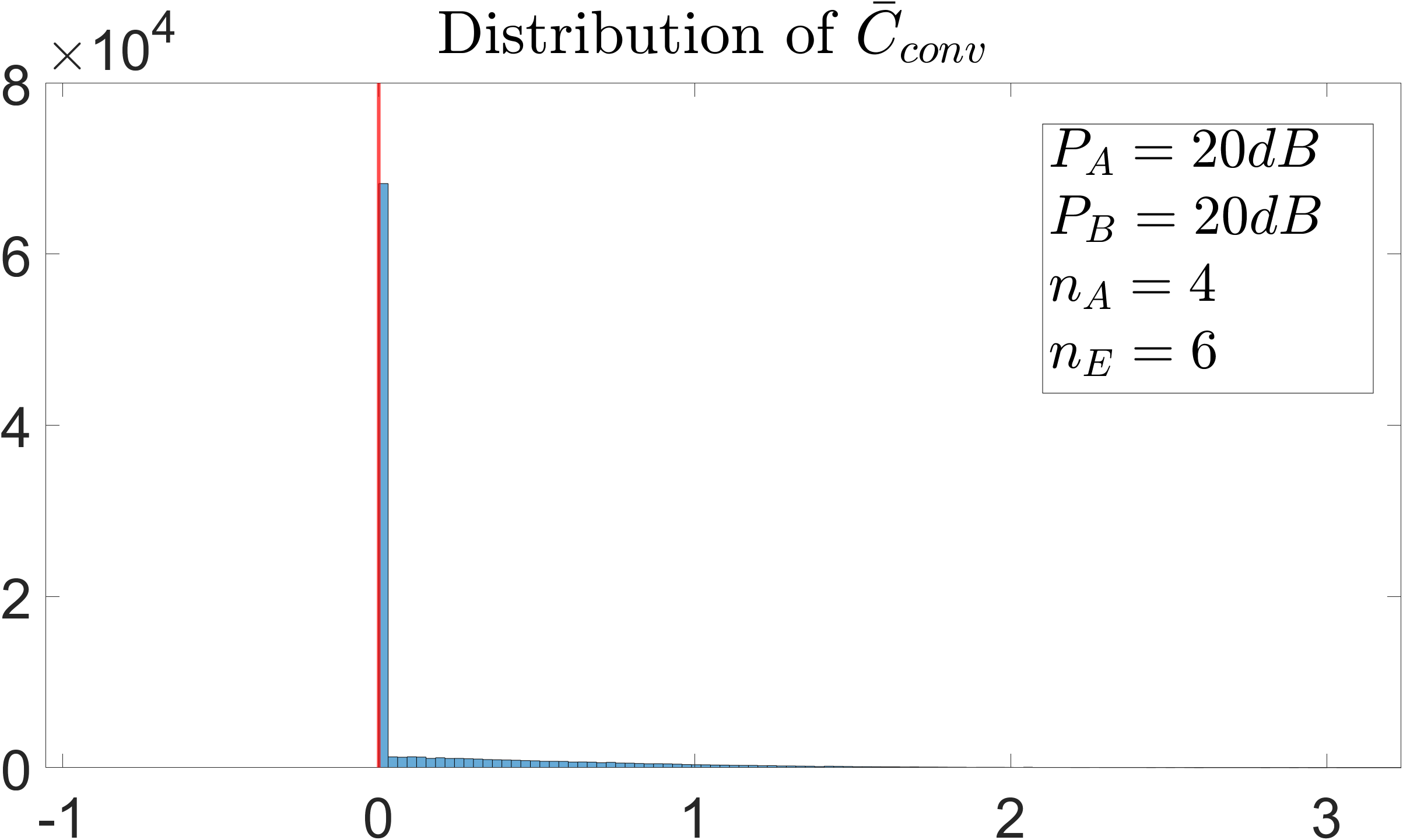}
\subcaption{$n_E=6$ and $P_B=20$dB.}
\label{fig:bar_C_conv_c}
\end{minipage}
\hspace{0.1cm}
\begin{minipage}[b]{0.45\linewidth}
\centering
\includegraphics[width=\textwidth]{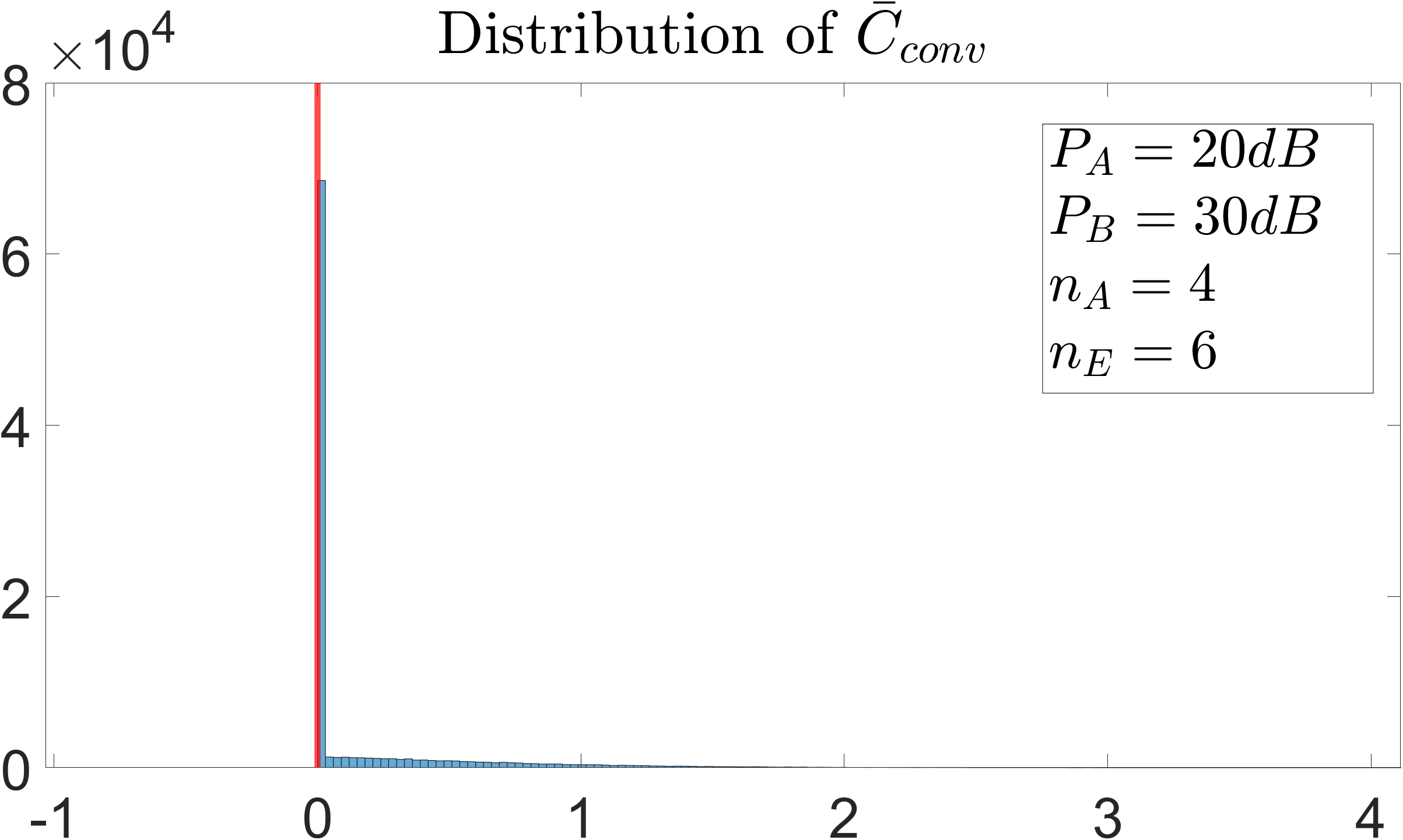}
\subcaption{$n_E=6$ and $P_B=30$dB.}
\label{fig:bar_C_conv_d}
\end{minipage}
\caption{Distribution of $\bar C_{\texttt{conv}}$ for $n_A=4$ and $P_A=20$dB and under the same conditions as for Fig. \ref{fig:bar_C_steep}.}
\label{fig:bar_C_conv}
\end{figure}

Under the exactly same conditions as for Fig. \ref{fig:bar_C_steep}, Fig. \ref{fig:bar_C_conv} shows the distributions of $\bar C_{\texttt{conv}}$. The upper two plots in Fig. \ref{fig:bar_C_conv} are for $n_E=2$, which show a significant probability of natural outage (i.e., $O_{\texttt{conv}}(0)=\texttt{Prob}(\bar C_{\texttt{conv}}=0)$). The lower two plots in Fig. \ref{fig:bar_C_conv} are for $n_E=6$, which show the natural outage reaching near 100\%.

\begin{figure}[ht]
\begin{minipage}[b]{0.45\linewidth}
\centering
\includegraphics[width=\textwidth]{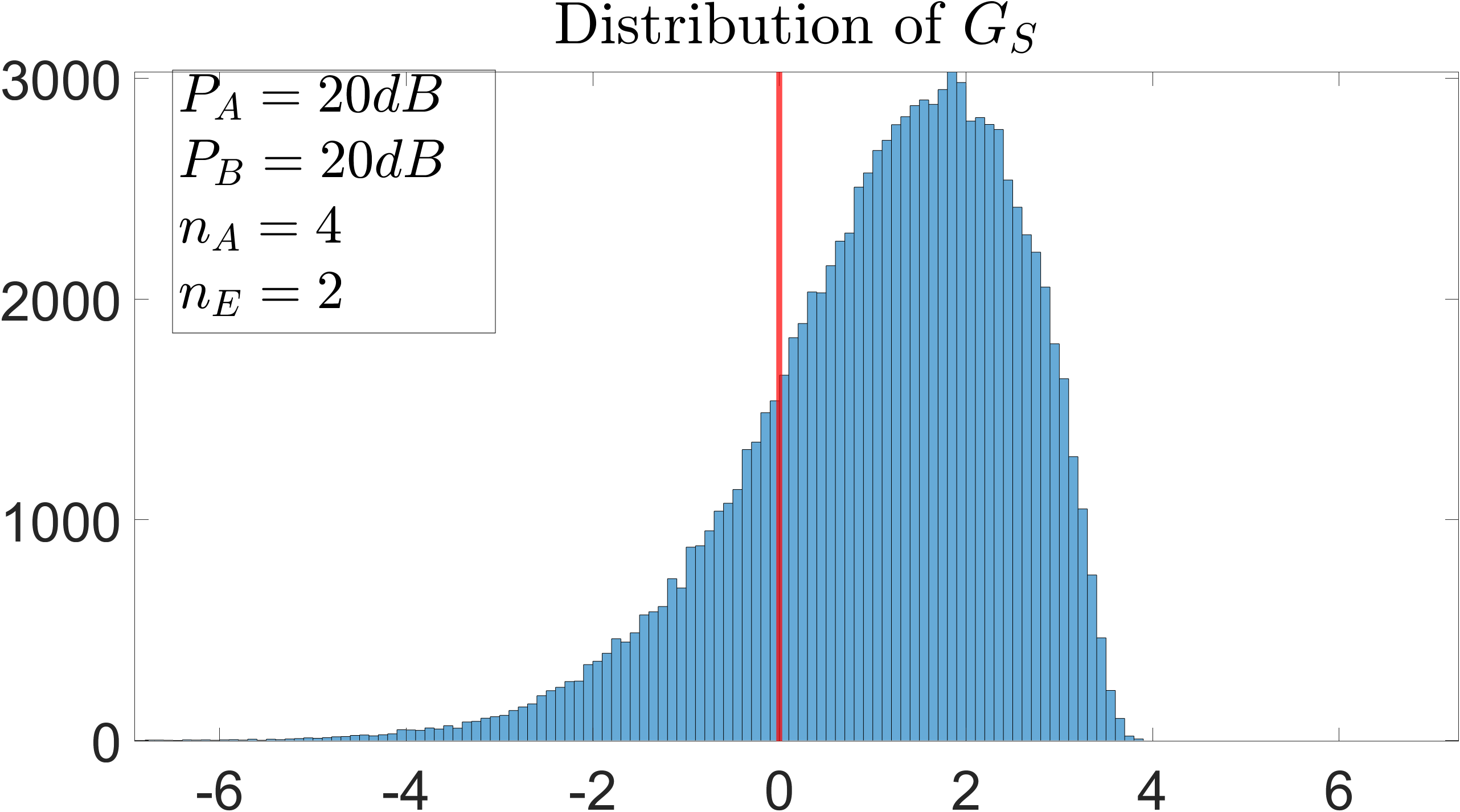}
\subcaption{$n_E=2$ and $P_B=20$dB.}
\label{fig:gain_a}
\end{minipage}
\hspace{0.1cm}
\begin{minipage}[b]{0.45\linewidth}
\centering
\includegraphics[width=\textwidth]{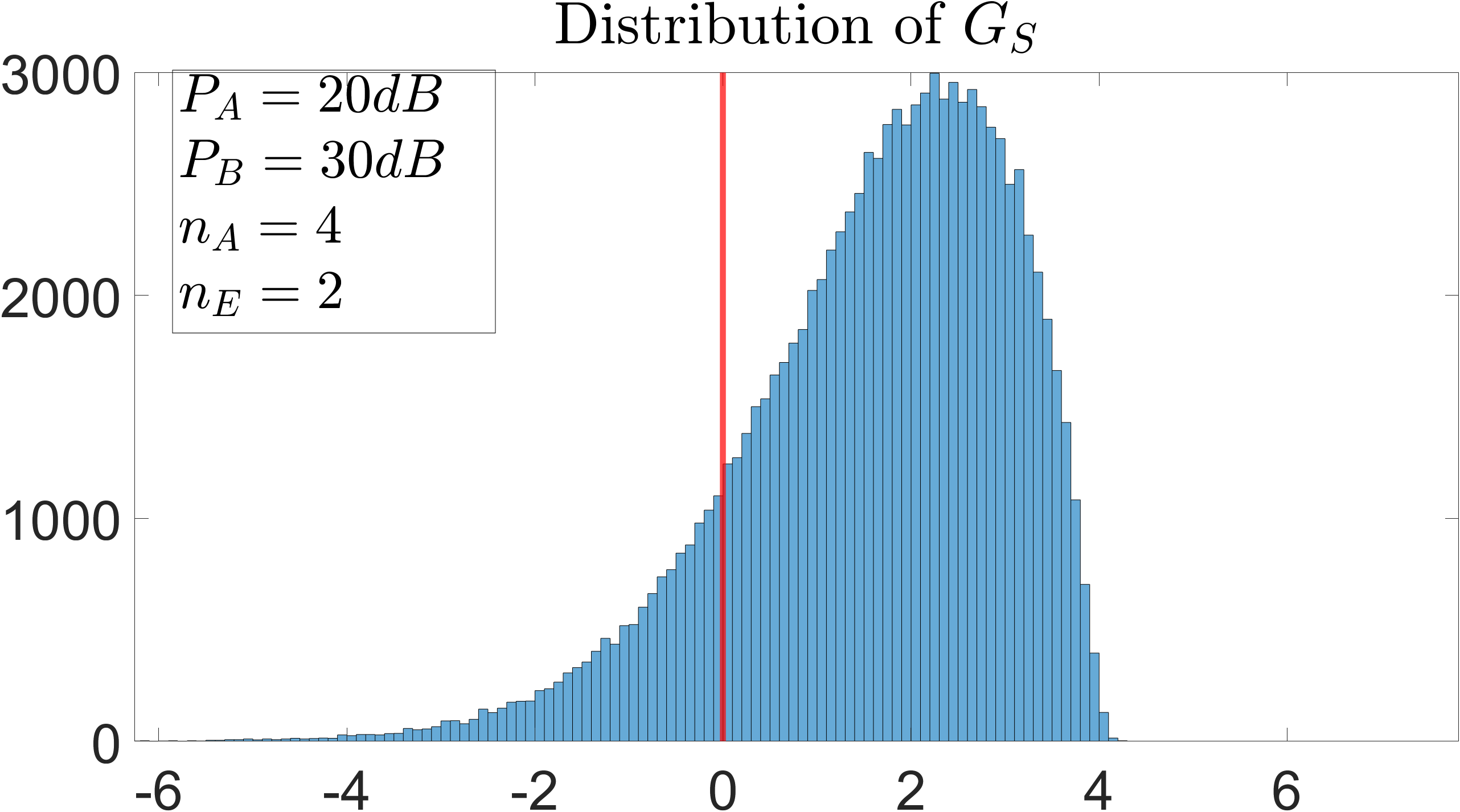}
\subcaption{$n_E=2$ and $P_B=30$dB.}
\label{fig:gain_b}
\end{minipage}
\\
\begin{minipage}[b]{0.45\linewidth}
\centering
\includegraphics[width=\textwidth]{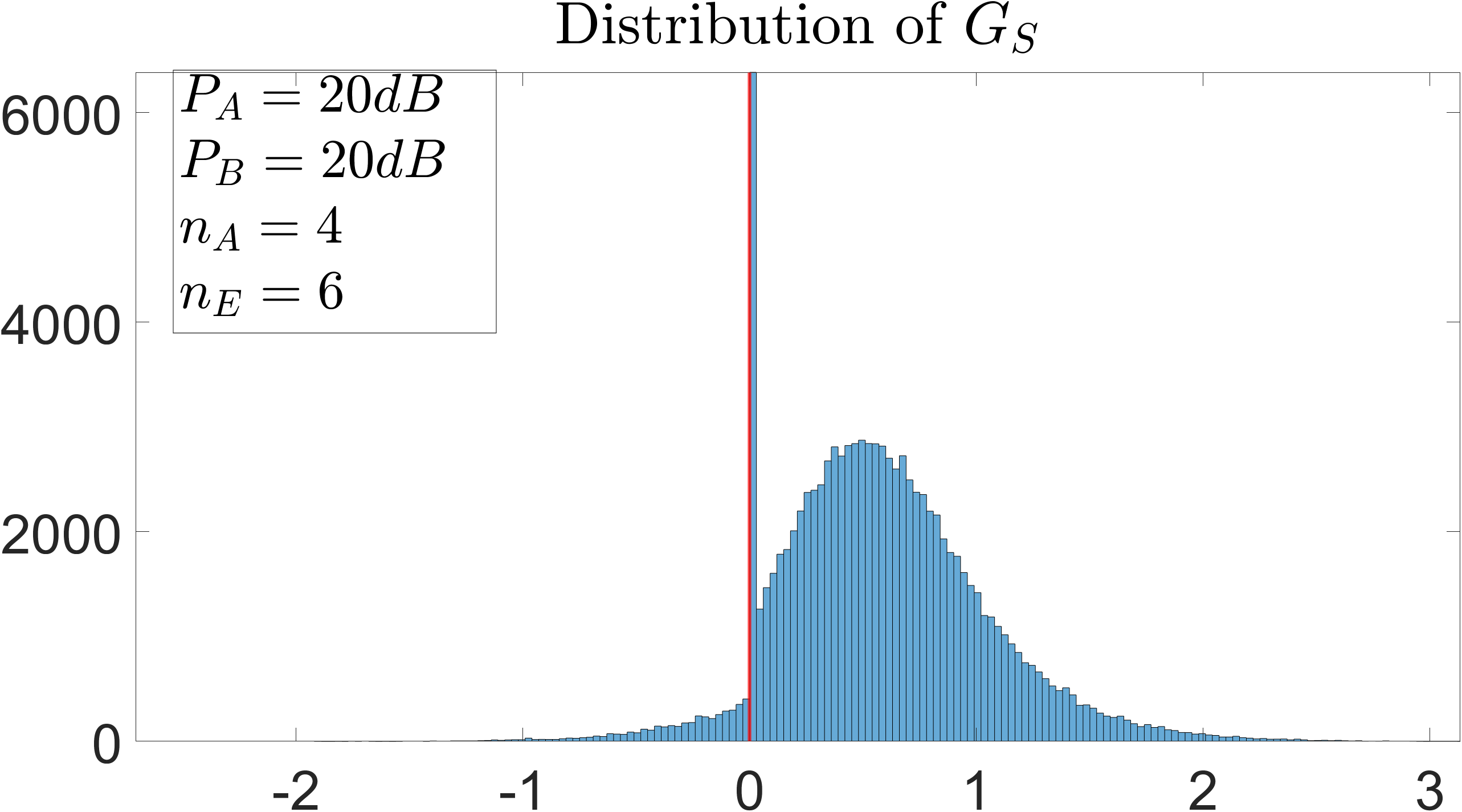}
\subcaption{$n_E=6$ and $P_B=20$dB.}
\label{fig:gain_c}
\end{minipage}
\hspace{0.1cm}
\begin{minipage}[b]{0.45\linewidth}
\centering
\includegraphics[width=\textwidth]{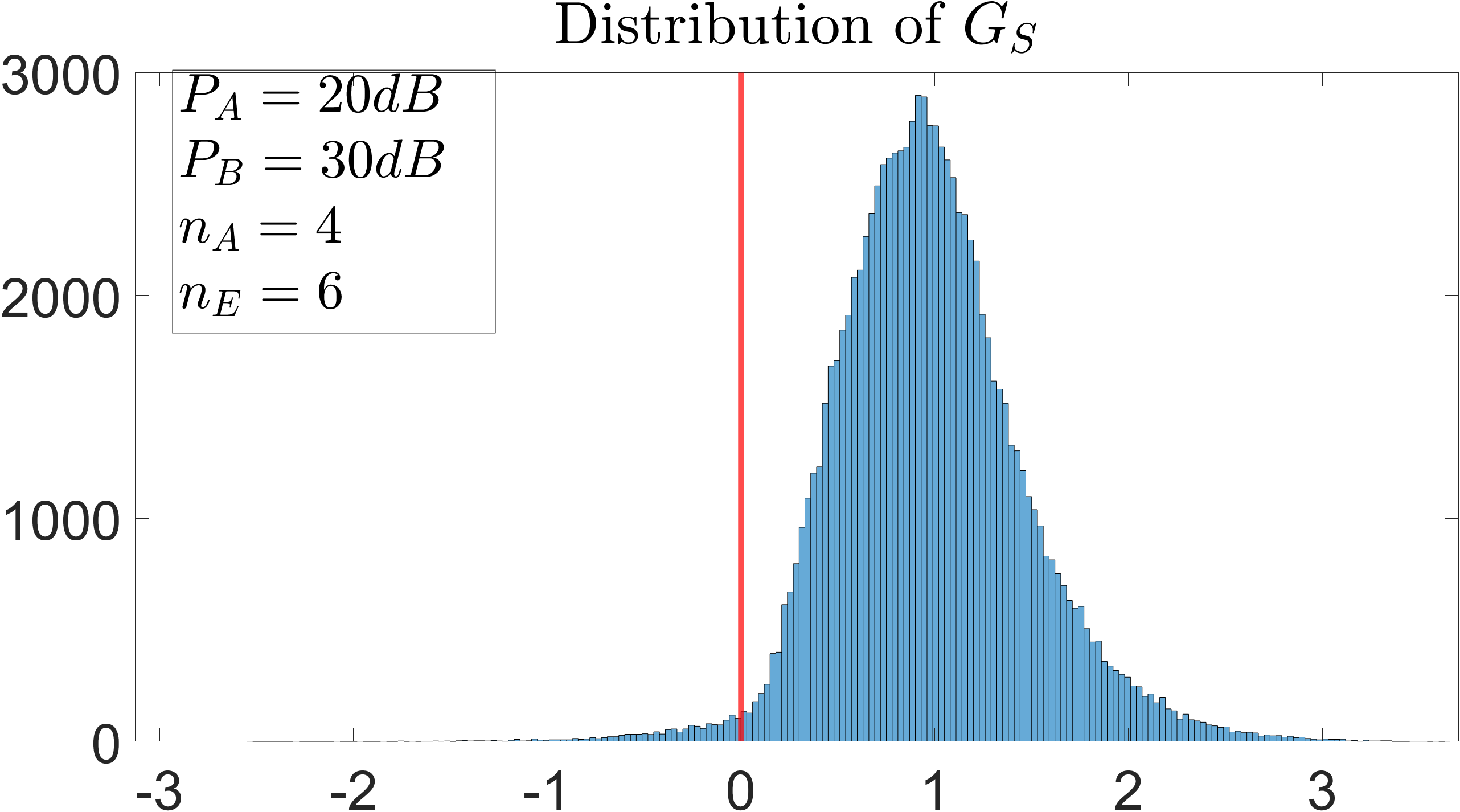}
\subcaption{$n_E=6$ and $P_B=30$dB.}
\label{fig:gain_d}
\end{minipage}
\caption{Distribution of $G_s$ for $n_A=4$ and $P_A=20$dB.}
\label{fig:gain}
\end{figure}

Fig. \ref{fig:gain} shows the distributions of the secrecy capacity gain $G_s$ of STEEP for $n_A=4$ and $P_A=20$dB. We see that this gain is mostly positive as expected from the previous analyses. This is true especially for a large $P_B$ and a large $n_E$. But we also see that there is a (small or not) probability that $G_s$ is negative. This is because when Eve's channel is in deep fade, the conventional scheme yields a positive secrecy in both directions of transmissions while STEEP does not take that advantage. However, this deep fade of Eve's channel has a low probability if Eve has a significant number of antennas. See Figs. \ref{fig:gain_c} and \ref{fig:gain_d} where $n_E=6$. Also, since Eve's channel is generally unknown to users, it seems infeasible for the conventional scheme to exploit Eve's deep fade.

\begin{figure}[ht]
\begin{minipage}[b]{0.45\linewidth}
\centering
\includegraphics[width=\textwidth]{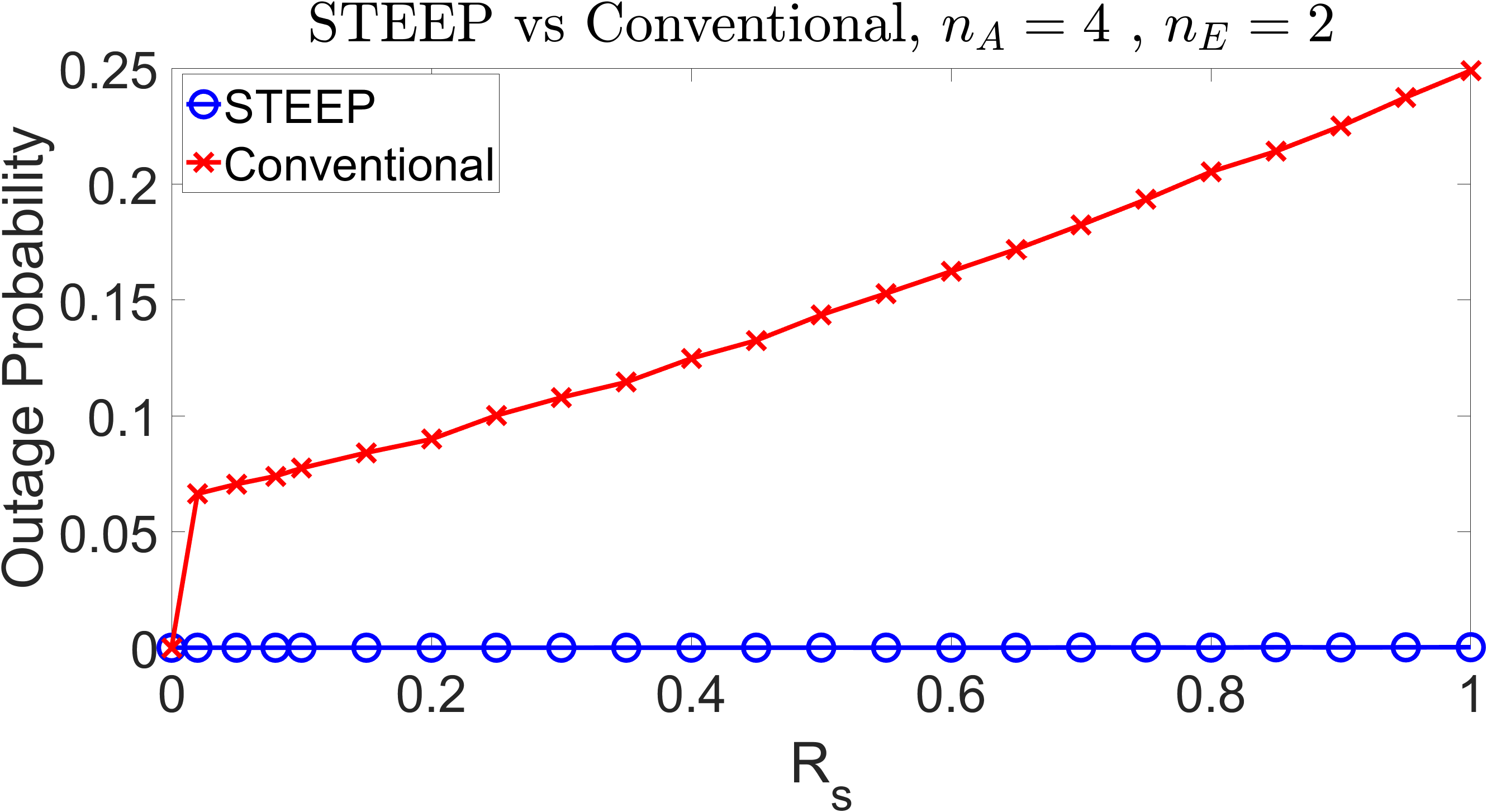}
\subcaption{$n_E=2$.}
\label{fig:outage_a}
\end{minipage}
\hspace{0.1cm}
\begin{minipage}[b]{0.45\linewidth}
\centering
\includegraphics[width=\textwidth]{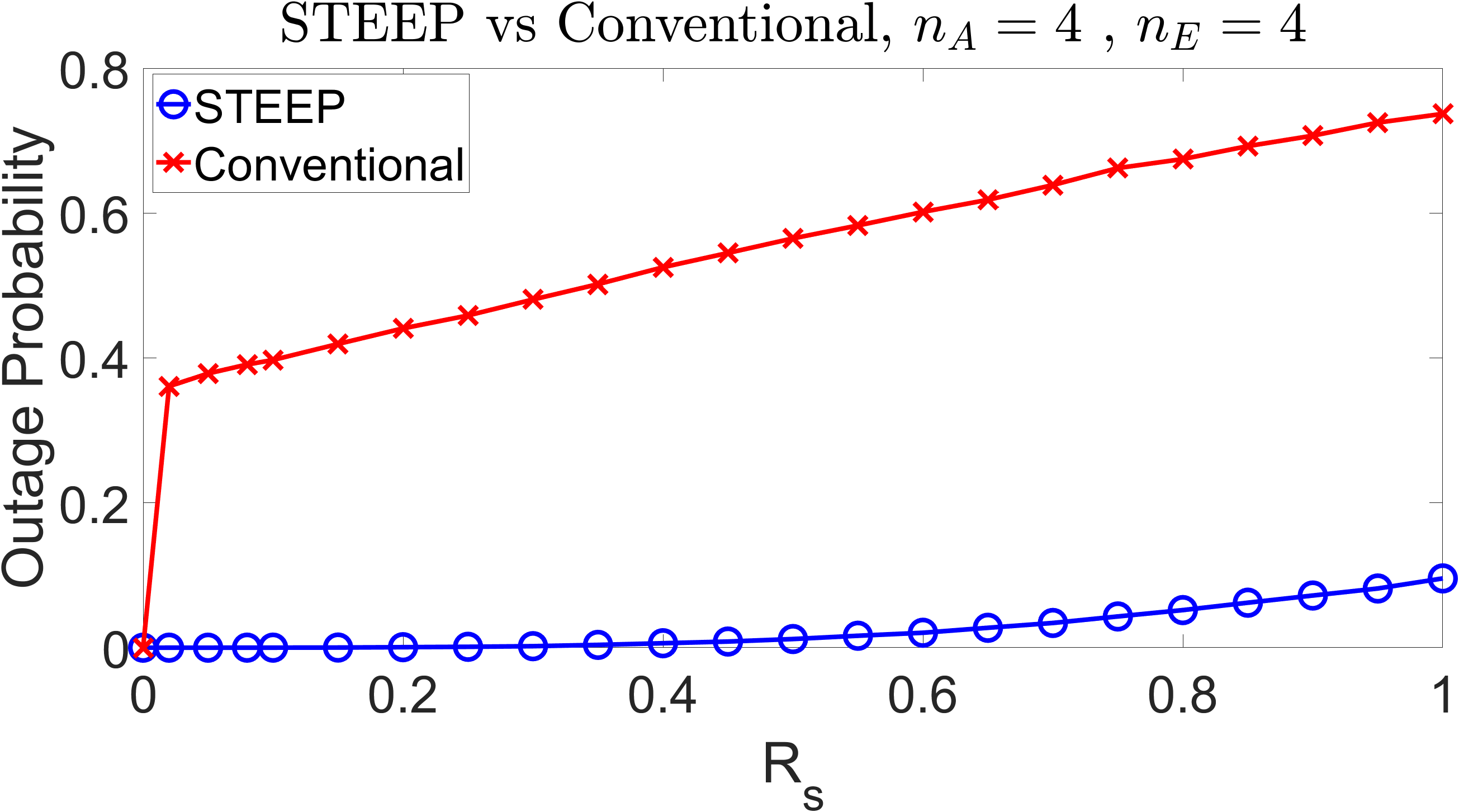}
\subcaption{$n_E=4$.}
\label{fig:outage_b}
\end{minipage}
\\\\
\begin{minipage}[b]{0.45\linewidth}
\centering
\includegraphics[width=\textwidth]{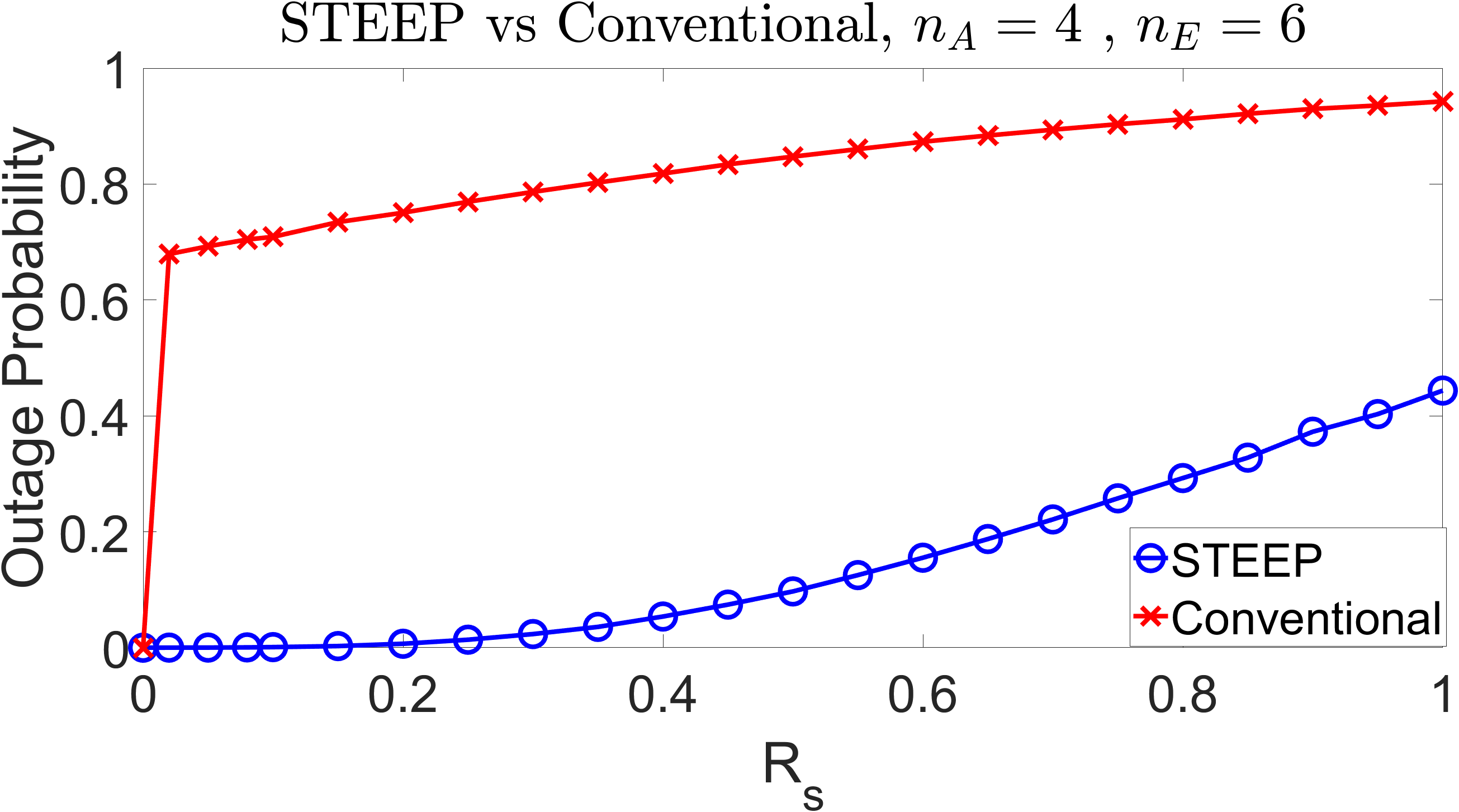}
\subcaption{$n_E=6$.}
\label{fig:outage_c}
\end{minipage}
\hspace{0.1cm}
\begin{minipage}[b]{0.45\linewidth}
\centering
\includegraphics[width=\textwidth]{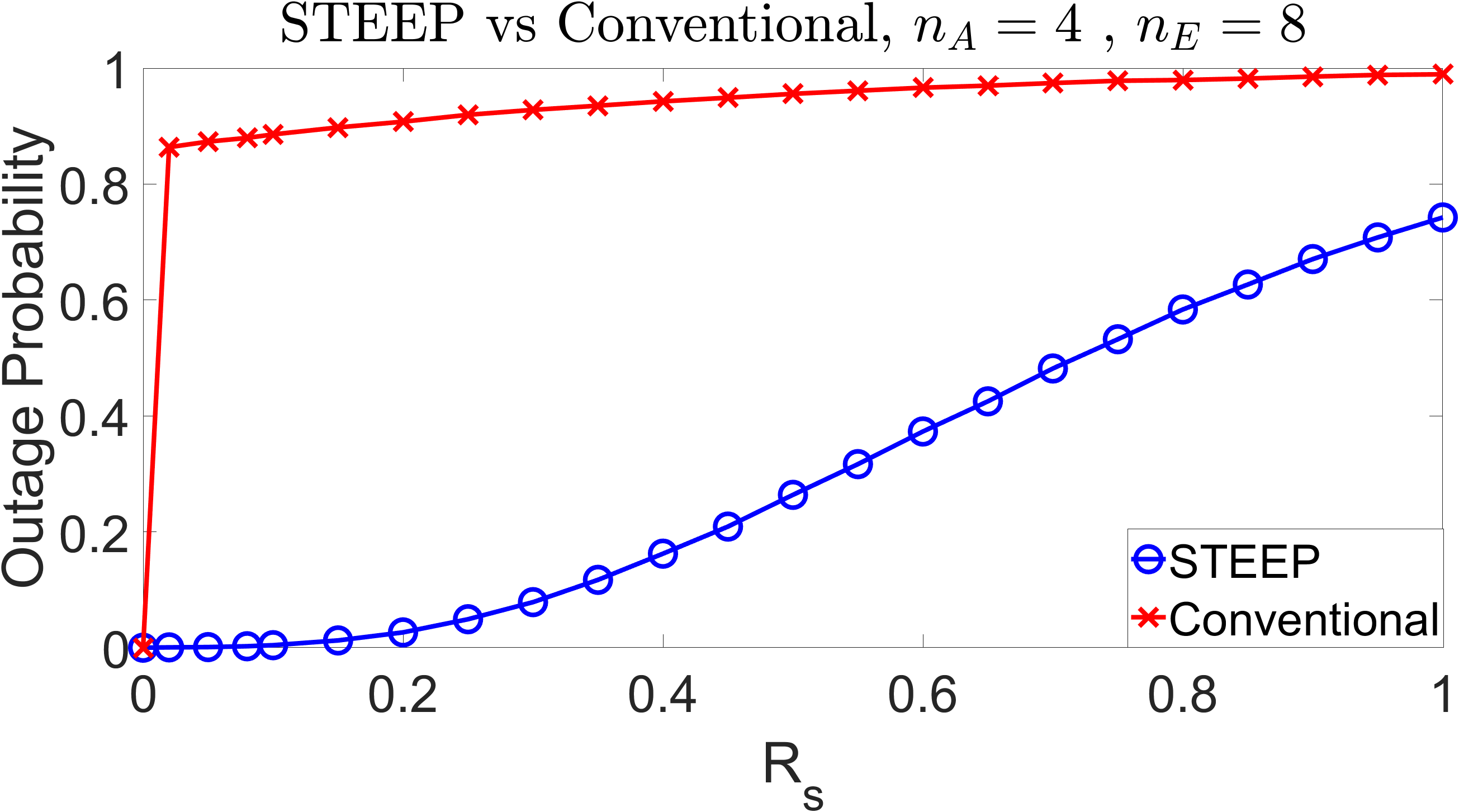}
\subcaption{$n_E=8$.}
\label{fig:outage_d}
\end{minipage}
\caption{Outage probabilities versus $R_s$ for $n_A=4$, $P_A=20$dB and $P_B=30$dB.}
\label{fig:outage}
\end{figure}

Fig. \ref{fig:outage} compares the outage probabilities of STEEP and the conventional as functions of the target secrecy rate $R_s$ (i.e., $O_{\texttt{STEEP}}(R_s)$ and $O_{\texttt{conv}}(R_s)$) where $n_A=4$, $P_A=20$dB and $P_B=30$dB. We see that for all the cases of $n_E=2,4,6,8$ and $0<R_s<1$, STEEP has much smaller outage probabilities than the conventional scheme.

\begin{figure}[ht]
\begin{minipage}[b]{0.45\linewidth}
\centering
\includegraphics[width=\textwidth]{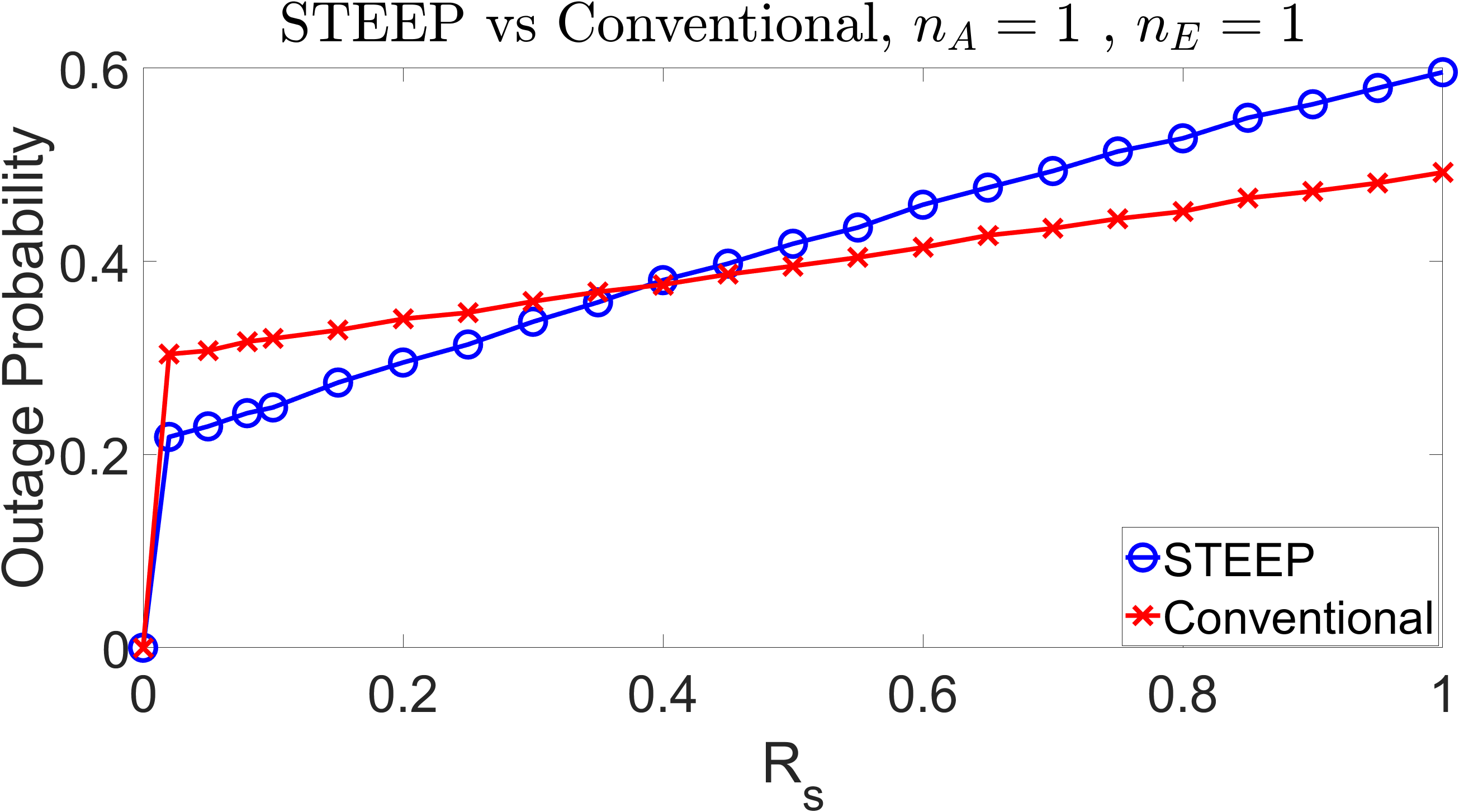}
\subcaption{$n_E=1$ and $P_B=30$dB.}
\label{fig:outage_nA_1_a}
\end{minipage}
\hspace{0.1cm}
\begin{minipage}[b]{0.45\linewidth}
\centering
\includegraphics[width=\textwidth]{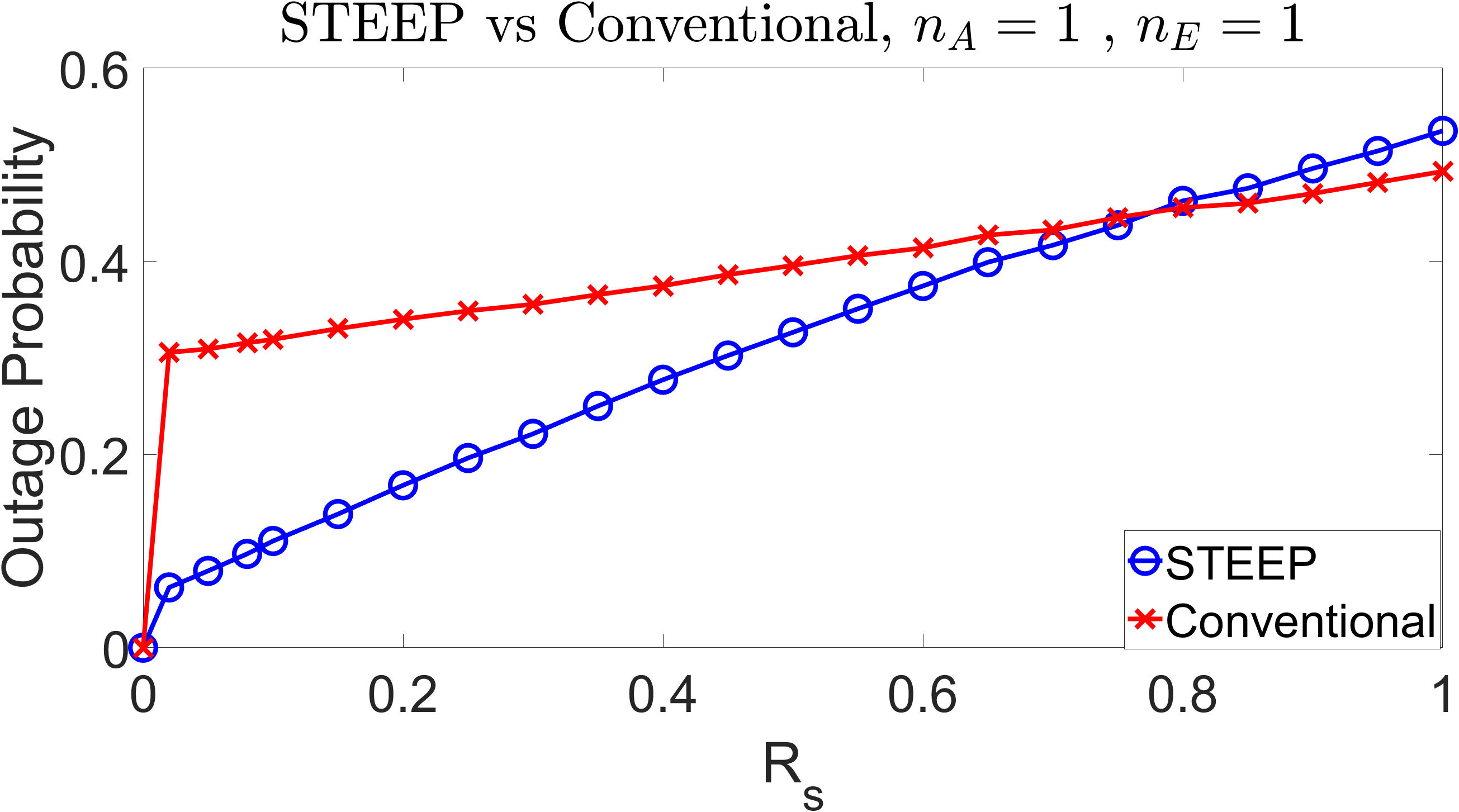}
\subcaption{$n_E=1$ and $P_B=40$dB.}
\label{fig:outage_nA_1_b}
\end{minipage}
\\
\begin{minipage}[b]{0.45\linewidth}
\centering
\includegraphics[width=\textwidth]{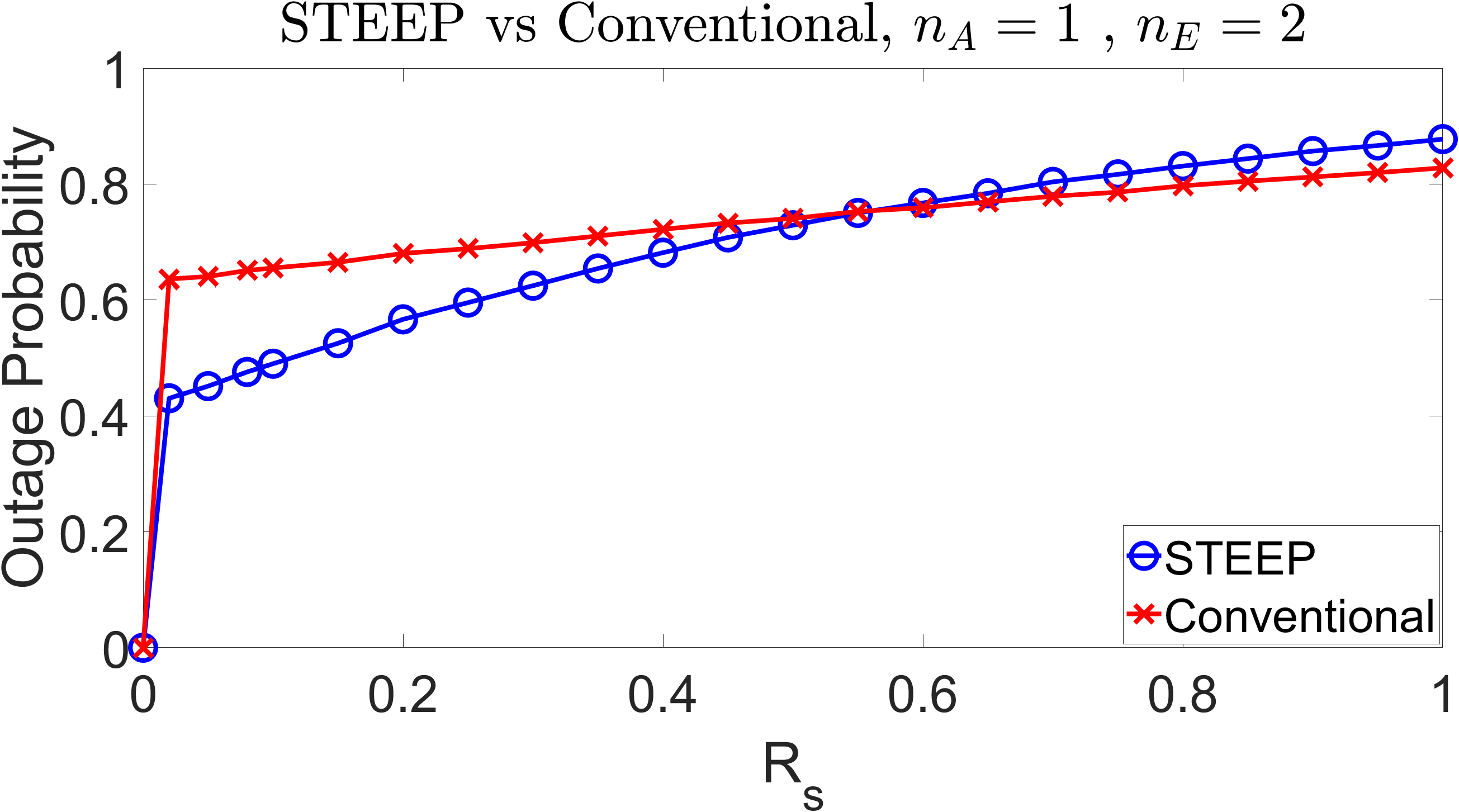}
\subcaption{$n_E=2$ and $P_B=30$dB.}
\label{fig:outage_nA_1_c}
\end{minipage}
\hspace{0.1cm}
\begin{minipage}[b]{0.45\linewidth}
\centering
\includegraphics[width=\textwidth]{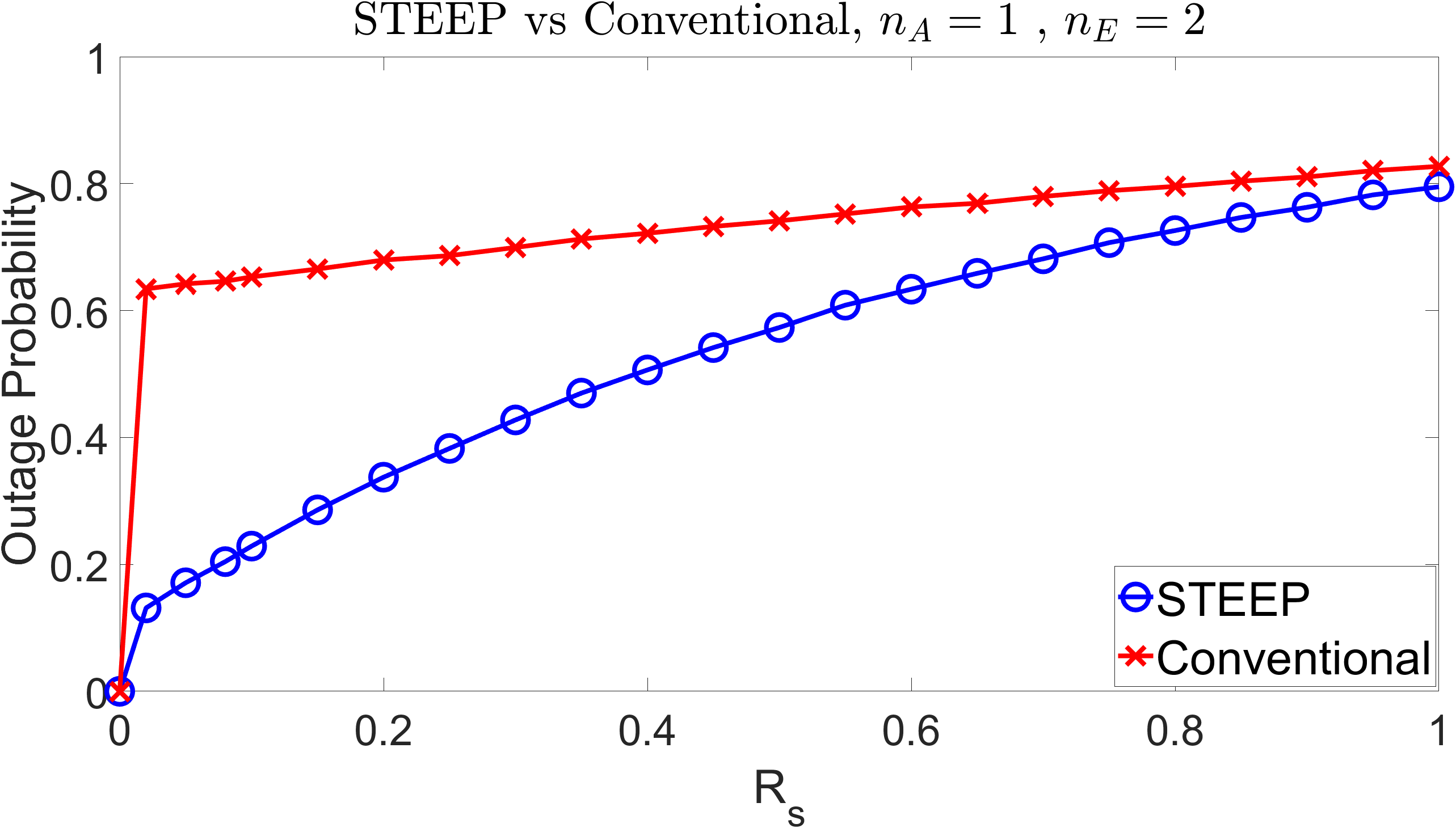}
\subcaption{$n_E=2$ and $P_B=40$dB..}
\label{fig:outage_nA_1_d}
\end{minipage}
\caption{Outage probabilities versus $R_s$ for $n_A=1$ and $P_A=20$dB.}
\label{fig:outage_nA_1}
\end{figure}

Finally, we show the case of $n_A=1$ in Fig. \ref{fig:outage_nA_1} with $P_A=20$dB. In this case, the channel between Alice and Bob is highly vulnerable due to fading, and there is no antenna diversity. Because of that, we see from Fig. \ref{fig:outage_nA_1_a} that STEEP and the conventional scheme do not have a large difference in terms of outage performance when $P_B$ is 30dB. But with $P_B=40$dB, the gap of outage performances in the lower end of $R_s$ becomes significantly larger (see Fig. \ref{fig:outage_nA_1_b}). A similar trend can be seen in Figs. \ref{fig:outage_nA_1_c} and \ref{fig:outage_nA_1_d} where $n_E=2$.

\section{Conclusion}
Further insights into STEEP have been provided. The secrecy capacity of STEEP is once again shown to be robust against the strength of Eve's channels, which opens a new door for secure communications. A more comprehensive analysis of STEEP is available in an upcoming paper.

\end{document}